%% file: main.tex
\newcommand{\AUTHORS}{Jonathan Frankle \\[.2cm] \emph{Advised by Professor David Walker}}
\newcommand{\TITLE}{Type-Directed Synthesis of Products}
\newcommand{\SUBTITLE}{Princeton University Master's Thesi-s}

\newcommand{\PAGENUMBERS}{yes}       

\documentclass[pdftex,twoside,twocolumn,12pt,letterpaper]{article}
\usepackage{ifthen}

\ifthenelse{\equal{\PAGENUMBERS}{yes}}{%
\usepackage[nohead,
            left=1in,right=1in,top=1in,
            footskip=0.5in,bottom=0.75in     
            ]{geometry}
}{%
\usepackage[noheadfoot,columnsep=0.2in,
            margin=1in,centering,truedimen]{geometry}
}

\usepackage{listings}
\usepackage{setspace}
\usepackage{wrapfig}
\usepackage{fancyhdr}
\usepackage[numbers,sort]{natbib}
\usepackage{xspace}
\usepackage{booktabs}
\usepackage{subcaption}
\usepackage[T1]{fontenc}
\usepackage{textcomp}
\usepackage{mathptmx}   
\usepackage{courier}
\usepackage{enumitem}
\usepackage{pbox}
\usepackage{verbatim}
\usepackage{changepage}
\usepackage{amsthm}
\usepackage{color}
\usepackage[usenames,dvipsnames]{xcolor}
\setlist{nolistsep}
\usepackage[scaled=0.92]{helvet}
\usepackage{fixltx2e}
\usepackage{bussproofs}

\usepackage{url}

\usepackage[pdftex]{graphicx}

\ifthenelse{\equal{\PAGENUMBERS}{yes}}{%
  \pagestyle{plain}
}{%
  \pagestyle{empty}
}

\makeatletter\long\def\@makecaption#1#2{
   \vskip 10pt
   \setbox\@tempboxa\hbox{\textsf{#1: #2}}
   \ifdim \wd\@tempboxa >\hsize 
       \textsf{#1: #2}\par      
     \else                      
       \hbox to\hsize{\hfil\box\@tempboxa\hfil}
   \fi}
\makeatother

\title{\textbf{\TITLE} \\[.4cm] \large \SUBTITLE}
\author{\AUTHORS}
\date{}

\def\compactify{\itemsep=0pt \topsep=0pt \partopsep=0pt \parsep=0pt} 
\let\latexusecounter=\usecounter

\newcommand{\ignore}[1]{}

\input{tds-macros}

\newcommand{\hlm}[1]{\setlength{\fboxsep}{1pt}\colorbox{CornflowerBlue!30}{$\displaystyle #1$}}
\newcommand{\hl} [1]{\setlength{\fboxsep}{1pt}\colorbox{CornflowerBlue!30}{#1}}
\DeclareMathAlphabet{\mathcal}{OMS}{cmsy}{m}{n}

\renewcommand{\TirNameStyle}[1]{\footnotesize \textsc{#1}}
\renewcommand{\TirName}[1]{\hbox {\small \TirNameStyle{#1}}}

\newcommand{\OneSp}{.3cm}
\newcommand{\TwoSp}{1.5cm}

\begin{document}
\clearpage\thispagestyle{empty}
\onecolumn
	\begin{doublespacing}
	\parbox[c]{6.5in}{\begin{center}\bfseries
    TYPE DIRECTED SYNTHESIS OF PRODUCTS\\\vspace{\TwoSp}
    Jonathan Frankle\\\vspace{\TwoSp} A THESIS\\\vspace{\OneSp}
	PRESENTED TO THE FACULTY\\\vspace{\OneSp}
	OF PRINCETON UNIVERSITY \\\vspace{\OneSp}
    IN CANDIDACY FOR THE DEGREE\\\vspace{\OneSp}
    OF MASTER OF SCIENCE IN ENGINEERING\\\vspace{\TwoSp}
    RECOMMENDED FOR ACCEPTANCE\\\vspace{\OneSp}
    BY THE DEPARTMENT OF\\\vspace{\OneSp}
    COMPUTER SCIENCE\\\vspace{\OneSp}
    Adviser: David Walker\\\vspace{\TwoSp}
    September 2015\par
	\end{center}
}
\end{doublespacing}
\newpage
\clearpage\thispagestyle{empty}
\vspace*{3cm}
\begin{center}
\copyright~Copyright by Jonathan Frankle, 2015.  All rights reserved.
\end{center}
\newpage

\onecolumn
\normalsize
\input{abstract}   
\input{intro}
\input{tds}
\input{theory}
\input{evaluation}
\input{related}
\input{future-work}
\input{conclusions}
\input{acknowledgements}

\small
\bibliographystyle{abbrvnat}
\bibliography{local}

\newpage
\onecolumn
\appendix
\input{proofs}

\end{document}

%% file: tds-macros.tex
\usepackage{mathpartir}
\usepackage{amsfonts}
\usepackage{amsmath}
\usepackage{xparse}
\usepackage{stmaryrd}

\newcommand{\CtorCtx} {\ensuremath{\Sigma}}      
\newcommand{\TypeCtx} {\ensuremath{\Gamma}}      
\newcommand{\ValCtx}  {\ensuremath{\Delta}}      
\newcommand{\ExamCtx} {\ensuremath{\mathcal{E}}} 
\newcommand{\FocusCtx}{\ensuremath{\mathcal{F}}} 
\newcommand{\AuxCtx}  {\ensuremath{\mathcal{A}}} 
\newcommand{\GoalCtx} {\ensuremath{\mathcal{G}}} 
\newcommand{\WorldOne}{\ensuremath{w}}           
\newcommand{\WorldCtx}{\ensuremath{\mathcal{W}}} 
\newcommand{\Exam}{\ensuremath{\epsilon}}        
\newcommand{\Refinement}{\ensuremath{r}}         
\newcommand{\Partial}{\ensuremath{pf}}           
\newcommand{\Elim}{\ensuremath{E}}               
\newcommand{\Intro}{\ensuremath{I}}              

\newcommand{\Wf}[1]{\ensuremath{\vdash {#1}~~wf}} 

\newcommand{\DType}{\ensuremath{\tau}}           
\newcommand{\DVar} {\ensuremath{x}}              
\newcommand{\DExp} {\ensuremath{e}}              
\newcommand{\DVal} {\ensuremath{v}}              
\newcommand{\DValEq}{\ensuremath{\dot{\DVal}}}   
\newcommand{\DExam}{\ensuremath{\Exam}}          
\newcommand{\DCtor}{\ensuremath{C}}              
\newcommand{\DFun} {\ensuremath{f}}              

\newcommand{\GBar}{\ensuremath{~|~}}

\newcommand{\GEq}{\ensuremath{::=~}}
\newcommand{\GEmp}{\ensuremath{\cdot}}

\DeclareDocumentCommand{\Range}{                
    O{\DVal} O{n} O{,}
  }{\ensuremath{{#1}_1{#3}~... {#3}~{#1}_{#2}}}
\newcommand{\CtorType}[1]                       
   {{#1} \rightarrow \TBase}
\newcommand{\Overbar}[2]                        
  {\ensuremath{\overline{{#2}}^{#1}}}
\newcommand{\SynthProp}[4]                      
   {\ensuremath{{#2} \vdash {#3} \overset{#1}{\leadsto} {#4}}}
\newcommand{\IntSynthProp}[3]                   
   {\ensuremath{{#2} \overset{#1}{\leadsto} {#3}}}

\DeclareDocumentCommand{\HasType}{               
    O{\DExp} O{\DType}
  }{\ensuremath{{#1} : {#2}}}
\DeclareDocumentCommand{\HasTypeCtx}{               
    O{\TypeCtx} O{\DExp} O{\DType}
  }{\ensuremath{{#1} \vdash {#2} : {#3}}}
\DeclareDocumentCommand{\HasTypeVal}{               
    O{\DExp} O{\DType} O{\DVal}
  }{\ensuremath{{#1} : {#2} = {#3}}}
\DeclareDocumentCommand{\InType}{                
    O{\DExp} O{\DType}
  }{\ensuremath{{#1} \in {#2}}}
\DeclareDocumentCommand{\Subtype}{               
    O{\DType_1} O{\DType_2}
  }{\ensuremath{{#1} <: {#2}}}
\DeclareDocumentCommand{\Refines}{               
    O{\DVar} O{\DExam}
  }{\ensuremath{{#1} \triangleright {#2}}}
\DeclareDocumentCommand{\SubRefines}{            
    O{\Refinement_1} O{\Refinement_2} O{\DType}
  }{\ensuremath{\Refines[\Subtype[{#1}][{#2}]][{#3}]}}

\newcommand{\TBase}{\ensuremath{\mathcal{B}}}   
\DeclareDocumentCommand{\TComp}{                
    O{\TBase} O{\DVal}
  }{\ensuremath{{#1}^{\setminus {#2}}}}
\DeclareDocumentCommand{\TAnd}{                 
    O{\DType_1} O{\DType_2}
  }{\ensuremath{{#1} \land {#2}}}
\DeclareDocumentCommand{\TAndRange}{            
    O{\DType} O{m}
  }{\Range[#1][#2][~\land]}
\DeclareDocumentCommand{\TAnds}{                
    O{\DType_1}
  }{\ensuremath{\land({#1})}}
\DeclareDocumentCommand{\TAndsRange}{           
    O{\DType} O{m}
  }{\TAnds[{#1}_1, ..., {#1}_{#2}]}
\DeclareDocumentCommand{\TOr}{                 
    O{\DType_1} O{\DType_2}
  }{\ensuremath{{#1} \lor {#2}}}
\DeclareDocumentCommand{\TOrRange}{            
    O{\DType} O{m}
  }{\Range[#1][#2][~\lor]}
\DeclareDocumentCommand{\TOrs}{                
    O{\DType_1}
  }{\ensuremath{\lor({#1})}}
\DeclareDocumentCommand{\TOrsRange}{           
    O{\DType} O{m}
  }{\TOrs[{#1}_1, ..., {#1}_{#2}]}
\newcommand{\TStar}{\ensuremath{*}}             
\newcommand{\TTuple}[1]{\ensuremath{#1}}        
\DeclareDocumentCommand{\TTupleRange}{          
    O{\DType} O{m}
  }{\Range[#1][#2][~\TStar]}
\newcommand{\TFunc}[2]                          
   {\ensuremath{{#1} \rightarrow {#2}}}
\newcommand{\TFuncE}[2]                         
   {\ensuremath{{#1} \Rightarrow {#2}}}
\newcommand{\DTFunc}{\TFunc{\DType_1}{\DType_2}}

\DeclareDocumentCommand{\TVal}{                 
    O{\DValEq}
  }{\ensuremath{\langle {#1} \rangle}}
\newcommand{\TNot}[1]{\ensuremath{\textsf{not}({#1})}} 

\DeclareDocumentCommand{\RefineCtxO             
  }{O{\CtorCtx} O{\TypeCtx} O{\WorldCtx}
  }{\ensuremath{{#1}~|~{#2}~|~{#3}}}
\DeclareDocumentCommand{\EGuessCtxO             
  }{O{\CtorCtx} O{\TypeCtx}
  }{\ensuremath{{#1}~|~{#2}}}
\DeclareDocumentCommand{\EGuessO                
  }{O{\CtorCtx} O{\TypeCtx} O{\DType} O{\Intro}
  }{\SynthProp{\Elim}{\EGuessCtxO[#1][#2]}{#3}{#4}}
\DeclareDocumentCommand{\RefineO                
  }{O{\CtorCtx} O{\TypeCtx} O{\WorldCtx} O{\DType} O{\Intro}
  }{\SynthProp{\Intro}{\RefineCtxO[#1][#2][#3]}{#4}{#5}}

\DeclareDocumentCommand{\RefineCtxOM             
  }{O{\TypeCtx} O{\WorldCtx}
  }{\ensuremath{{#1}~|~{#2}}}
\DeclareDocumentCommand{\EGuessCtxOM             
  }{O{\TypeCtx}
  }{\ensuremath{{#1}}}
\DeclareDocumentCommand{\EGuessOM                
  }{O{\TypeCtx} O{\DType} O{\Intro}
  }{\SynthProp{\Elim}{\EGuessCtxOM[#1]}{#2}{#3}}
\DeclareDocumentCommand{\RefineOM                
  }{O{\TypeCtx} O{\WorldCtx} O{\DType} O{\Intro}
  }{\SynthProp{\Intro}{\RefineCtxOM[#1][#2]}{#3}{#4}}
\DeclareDocumentCommand{\ERefineOM               
  }{O{\TypeCtx} O{\WorldCtx} O{\DType} O{\Intro}
  }{\SynthProp{\Elim}{\RefineCtxOM[#1][#2]}{#3}{#4}}
\DeclareDocumentCommand{\SequentOM               
  }{O{\TypeCtx} O{\WorldCtx} O{\DType} O{\DExp}
  }{\SynthProp{}{\RefineCtxOM[#1][#2]}{#3}{#4}}

\DeclareDocumentCommand{\RefineCtxSeq             
  }{O{\TypeCtx} O{\WorldCtx}
  }{\ensuremath{{#1}~|~{#2}}}
\DeclareDocumentCommand{\EGuessCtxOM             
  }{O{\TypeCtx}
  }{\ensuremath{{#1}}}
\DeclareDocumentCommand{\EGuessOM                
  }{O{\TypeCtx} O{\DType} O{\Intro}
  }{\SynthProp{\Elim}{\EGuessCtxOM[#1]}{#2}{#3}}
\DeclareDocumentCommand{\RefineOM                
  }{O{\TypeCtx} O{\WorldCtx} O{\DType} O{\Intro}
  }{\SynthProp{\Intro}{\RefineCtxOM[#1][#2]}{#3}{#4}}
\DeclareDocumentCommand{\ERefineOM               
  }{O{\TypeCtx} O{\WorldCtx} O{\DType} O{\Intro}
  }{\SynthProp{\Elim}{\RefineCtxOM[#1][#2]}{#3}{#4}}
\DeclareDocumentCommand{\SequentOM               
  }{O{\TypeCtx} O{\WorldCtx} O{\DType} O{\DExp}
  }{\SynthProp{}{\RefineCtxOM[#1][#2]}{#3}{#4}}

\DeclareDocumentCommand{\World                  
  }{O{\ExamCtx} O{\Exam}
  }{\ensuremath{\langle{#1}; {#2}\rangle}}
\DeclareDocumentCommand{\WorldN                 
  }{O{\ExamCtx} O{\Exam}
  }{\Overbar{n}{\ensuremath{\langle{#1}; {#2}\rangle}}}
\DeclareDocumentCommand{\RefineCtx              
  }{O{\CtorCtx} O{\TypeCtx} O{\WorldCtx} O{\AuxCtx} O{\FocusCtx}
  }{\ensuremath{{#1}~|~{#2}; {#3}; {#4}~|~{#5}}}
\DeclareDocumentCommand{\EGuessCtx              
  }{O{\CtorCtx} O{\TypeCtx} O{\AuxCtx} O{\FocusCtx}
  }{\ensuremath{{#1}~|~{#2}; {#3}; {#4}}}
\DeclareDocumentCommand{\EGuess                 
  }{O{\CtorCtx} O{\TypeCtx} O{\AuxCtx} O{\FocusCtx} O{\DType} O{\Intro}
  }{\SynthProp{\Elim}{\EGuessCtx[#1][#2][#3][#4]}{#5}{#6}}
\DeclareDocumentCommand{\Refine                 
  }{O{\CtorCtx} O{\TypeCtx} O{\AuxCtx} O{\FocusCtx} O{\WorldCtx} O{\DType} O{\Intro}
  }{\SynthProp{\Intro}{\RefineCtx[#1][#2][#3][#4][#5]}{#6}{#7}}

\DeclareDocumentCommand{\RefineCtxM             
  }{O{\TypeCtx} O{\WorldCtx} O{\AuxCtx} O{\FocusCtx}
  }{{#1}; {#2}; {#3}~|~{#4}}
\DeclareDocumentCommand{\RefineM                
  }{O{\TypeCtx} O{\AuxCtx} O{\FocusCtx} O{\WorldCtx} O{\DType} O{\Intro}
  }{\SynthProp{\Intro}{\RefineCtx[#1][#2][#3][#4]}{#5}{#6}}
\DeclareDocumentCommand{\EGuessCtxM             
  }{O{\TypeCtx} O{\AuxCtx} O{\FocusCtx}
  }{\ensuremath{#1}; {#2}; {#3}}
\DeclareDocumentCommand{\EGuessM                
  }{O{\TypeCtx} O{\AuxCtx} O{\FocusCtx} O{\DType} O{\Elim}
  }{\SynthProp{\Elim}{\EGuessCtxM[#1][#2][#3]}{#4}{#5}}
\DeclareDocumentCommand{\RefineM                
  }{O{\TypeCtx} O{\AuxCtx} O{\FocusCtx} O{\WorldCtx} O{\DType} O{\Intro}
  }{\SynthProp{\Intro}{\RefineCtxM[#1][#2][#3][#4]}{#5}{#6}}
\DeclareDocumentCommand{\EGuessN                
  }{O{\TypeCtx} O{\AuxCtx} O{\FocusCtx} O{\DType} O{\Elim}
  }{\SynthProp{\Elim'}{\EGuessCtxM[#1][#2][#3]}{#4}{#5}}
\DeclareDocumentCommand{\RefineN                
  }{O{\TypeCtx} O{\AuxCtx} O{\FocusCtx} O{\WorldCtx} O{\DType} O{\Intro}
  }{\SynthProp{\Intro'}{\RefineCtxM[#1][#2][#3][#4]}{#5}{#6}}

\DeclareDocumentCommand{\IFocus                
  }{O{\RefineCtx} O{\RefineCtx}
  }{\ensuremath{{#1} \Longrightarrow {#2}}}
\DeclareDocumentCommand{\IFocusStar            
  }{O{\RefineCtx} O{\RefineCtx}
  }{\ensuremath{{#1} \FocusStarArrow {#2}}}

\DeclareDocumentCommand{\IntCtx                
  }{O{\CtorCtx} O{\TypeCtx} O{\AuxCtxL} O{\FocusCtxL} O{\FocusCtxR} O{\GoalCtx}
  }{\ensuremath{{#1}~|~{#2}; {#3}; {#4}~|~{#5}; {#6}}}
\DeclareDocumentCommand{\IntProp               
  }{O{\Intro} O{\Intro} O{\GoalCtx} O{\FocusCtxL} O{\FocusCtxR} O{\CtorCtx} O{\TypeCtx} O{\AuxCtxL}
  }{\IntSynthProp{#1}{\IntCtx[#6][#7][#8][#4][#5][#3]}{#2}}
\DeclareDocumentCommand{\IntPropMany           
  }{O{\Intro} O{\Intro} O{n} O{\GoalCtx_i} O{\FocusCtxL_i} O{\FocusCtxR_i} O{\CtorCtx_i} O{\TypeCtx_i} O{\AuxCtxL_i}
  }{\IntSynthProp{#1}{\Overbar{i \in #3}{\IntCtx[#7][#8][#9][#5][#6][#4]}}{#2}}
\DeclareDocumentCommand{\IntCtxI               
  }{O{i}
  }{\IntCtx[\CtorCtx_{#1}][\TypeCtx_{#1}][\AuxCtxL_{#1}][\FocusCtxL_{#1}][\FocusCtxR_{#1}][\GoalCtx_{#1}]}  
\DeclareDocumentCommand{\SynthCtx              
  }{O{i}}{\ensuremath{\Psi_{#1}}}    

\DeclareDocumentCommand{\IntMinCtx             
  }{O{\ValCtx} O{\TypeCtx} O{\DType}
  }{\ensuremath{{#1}~|~{#2}~|~{#3}}}
\DeclareDocumentCommand{\IntMinProp            
  }{O{\Intro} O{\Intro} O{\DType} O{\TypeCtx} O{\ValCtx}
  }{\IntSynthProp{#1}{\IntMinCtx[#5][#4][#3]}{#2}}
\DeclareDocumentCommand{\IntMinPropMany        
  }{O{\Intro} O{\Intro} O{n} O{\DType_i} O{\TypeCtx_i} O{\ValCtx_i}
  }{\IntSynthProp{#1}{\Overbar{i \in #3}{\IntMinCtx[#6][#5][#4]}}{#2}}
\DeclareDocumentCommand{\IntMinCtxI            
  }{O{i}
  }{\IntMinCtx[\ValCtx_{#1}][\TypeCtx_{#1}][\DType_{#1}]}

\DeclareDocumentCommand{\TSum}{O{\DType_1} O{\DType_2}
  }{\ensuremath{{#1} + {#2}}}
\DeclareDocumentCommand{\EInl}{O{\DExp}}{\ensuremath{\mathsf{inl}~{#1}}}
\DeclareDocumentCommand{\EInr}{O{\DExp}}{\ensuremath{\mathsf{inr}~{#1}}}
\DeclareDocumentCommand{\EMatchS}{O{\DExp} O{\DExp} O{\DExp}
  }{\ensuremath{\mathsf{match}~{#1}~\mathsf{with}~\EInl[\DVar] \rightarrow {#2}~|~
                                                  \EInr[\DVar] \rightarrow {#3}}}

\DeclareDocumentCommand{\EFix                  
  }{O{f} O{x} O{\DType_1} O{\DType_2} O{\DExp}
  }{\ensuremath{\mathsf{fix}~{#1} \HasType[(\HasType[{#2}][{#3}])][{#4}] = {#5}}}
\DeclareDocumentCommand{\ELam
  }{O{x} O{\DType_1} O{\DExp}
  }{\ensuremath{\lambda{#1}{:}{#2}.{#3}}}
\DeclareDocumentCommand{\ECtor                 
  }{O{\DCtor} O{\DExp}
  }{{\ensuremath{#1}}~{\ensuremath{#2}}}
\DeclareDocumentCommand{\EApp                  
  }{O{\DExp_1} O{\DExp_2}
  }{\ensuremath{{#1}~{#2}}}
\DeclareDocumentCommand{\EMatch                
  }{O{\DExp} O{\Overbar{i \in m}{\DCtor_i~\DVar \rightarrow e_i}}
  }{\ensuremath{\mathsf{match}~{#1}~\mathsf{with}~{#2}}}
\DeclareDocumentCommand{\ETuple                
  }{O{\DExp_1, ..., \DExp_m}
  }{\ensuremath{({#1})}}
\DeclareDocumentCommand{\EProj                 
  }{O{k} O{\DExp}
  }{\ensuremath{\pi_{#1}~{#2}}}
\DeclareDocumentCommand{\EProjI                
  }{O{k} O{\DExp}
  }{\ensuremath{\rho_{#1}~{#2}}}

\newcommand{\ETupleRange}[1]                   
  {\ETuple[\Range[#1][m]]}

\DeclareDocumentCommand{\RefCon
  }{O{\TypeCtx} O{\DVal} O{\DType} O{\Refinement}
  }{\ensuremath{\mathsf{RefCon}({#1}; {#2}; {#3}) = {#4}}}

\DeclareDocumentCommand{\Univ
  }{O{\Refinement} O{\DType} O{\Refinement'}
  }{\ensuremath{\mathsf{Univ}(\Refines[{#1}][{#2}]) = {#3}}}

\newcommand{\Sss}{\ensuremath{\rightarrow}}
\newcommand{\SssStar}{\ensuremath{\rightarrow^*}}
\DeclareDocumentCommand{\EtaEq}{O{\DExp} O{\DType}
    }{\ensuremath{\left[{#1}\right]_{\eta}^{#2}}}

\DeclareDocumentCommand{\TupDe}{O{\DExp} O{\DType}
    }{\ensuremath{\left[{#1}\right]_{tup}^{#2}}}
\DeclareDocumentCommand{\Vars}{O{\TypeCtx}}{\ensuremath{\left[{#1}\right]_{vars}}}
\DeclareDocumentCommand{\TNeg}{O{\DType}}{\ensuremath{\left[{#1}\right]_{neg}}}
\DeclareDocumentCommand{\TExp}{O{\DType} O{\DType}}{\ensuremath{\left[{#1}\right]_{exp}^{#2}}}
\DeclareDocumentCommand{\Denot}{O{\Refinement}}{\ensuremath{\llbracket{#1}\rrbracket}}
\DeclareDocumentCommand{\Norm}{O{\Refinement}}{\ensuremath{\mathcal{N}({#1})}}

%% file: abstract.tex
\section*{Abstract}
Software synthesis - the process of generating complete, general-purpose programs from
specifications - has become a hot research topic in the past few years.
For decades the problem was thought to be insurmountable: the search space of possible
programs is far too massive to efficiently traverse.  Advances in efficient
constraint solving have overcome this barrier, enabling a new generation
of effective synthesis systems.  Most existing systems compile synthesis tasks down
to low-level SMT instances, sacrificing high-level semantic information while solving only
first-order problems (i.e., filling integer holes).  Recent work
takes an alternative approach, using the Curry-Howard isomorphism and
techniques from automated theorem
proving to construct higher-order programs with algebraic datatypes.

My thesis involved
extending this \emph{type-directed} synthesis engine to handle product types, which
required significant modifications to both the underlying theory and the tool itself.
Product types streamline other language features, eliminating variable-arity
constructors among other workarounds employed in the original synthesis system.  A form
of logical conjunction, products are \emph{invertible}, making it possible to equip
the synthesis system with an efficient theorem-proving technique called \emph{focusing}
that eliminates many of the nondeterministic choices inherent in proof search.
These theoretical enhancements informed a new version of the type-directed synthesis prototype
implementation, which remained performance-competitive with the original synthesizer.
A significant advantage of the type-directed synthesis framework is its
extensibility; this thesis is a roadmap for future such efforts to increase the expressive
power of the system.

%% file: intro.tex
\section{Introduction}

Since the advent of computer programming, the cycle of writing, testing, and
debugging code has remained a tedious and error-prone undertaking.  There is
no middle ground between a program that is correct and one that is not, so
software development demands often-frustrating levels of precision.  The task
of writing code is mechanical and repetitive, with boilerplate and common idioms consuming
ever more valuable developer-time.  Aside from the languages and platforms,
the process of software engineering has changed little over the past four decades.

The discipline of software synthesis responds to these deficiencies with a simple
question: if computers can automate so much of our everyday lives, why can they
not do the same for the task of developing software?  Outside of a few specialized
domains, this question -- until recently -- had an equally simple answer:
the search space of candidate programs is too large to explore efficiently.
As we increase the number of abstract syntax tree (AST) nodes that a program might require,
the search space explodes combinatorially.  Synthesizing even
small programs seems hopeless.

In the past decade, however, several research program analysis and synthesis systems have overcome
these barriers and developed into useful programming tools.  At the core of
Sketch~\cite{sketch}, Rosette~\cite{rosette1, rosette2}, and
Leon~\cite{leon-recursive} are efficient SAT and SMT-solvers.  These tools
automate many development tasks -- including test case generation,
verification, angelic nondeterminism, and synthesis~\cite{rosette1} --
by compiling programs into constraint-solving problems.

These synthesis techniques, while a major technological step forward, are
still quite limited.  Constraint-solving tasks are fundamentally first-order:
although they efficiently fill integer and boolean holes, they cannot
scale to higher-order programs.  Furthermore, in the process of compiling
synthesis problems into the low-level language of constraint-solvers,
these methods sacrifice high-level semantic and type information that might
guide the synthesis procedure through the search space more efficiently.

Based on these observations, recent work by Steve
Zdancewic and Peter-Michael Osera at the Univerity of Pennsylvania~\cite{tds} explores
a different approach to synthesis: theorem-proving.  The Curry-Howard
isomorphism permits us to treat the type of a desired program as a theorem whose
proof is the synthesized program.  Translating this idea into an algorithm,
we can search for the solution to the synthesis problem using
existing automated theorem-proving techniques.
Since many programs inhabit the same type, users also
provide input-output examples to better specify the desired function and
constrain the synthesizer's result.

This \emph{type-directed} approach scales to higher-order programs and preserves
the high-level program structures that guide the synthesis process.
The synthesis algorithm can be designed to search only for well-typed programs in
normal form, drastically reducing the search space of possible ASTs.  At the time
of writing, both of these features are unique to type-directed synthesis.

Not only is this technique efficient, but it also has the ability to scale
to language features that have, until now, remained beyond the reach of synthesis.
Many desirable features, like file input-output,
map to existing systems of logic that, when integrated
into the synthesizer, instantly enable it to generate the corresponding programs.
For example, one might imagine synthesizing effectful computation using monads
by performing proof search in lax logic~\cite{pfenning-modal}.  When these logical
constructs are kept orthogonal to one another, features can be added and
removed from the synthesis language in modular fashion.
This extensibility is one of the most noteworthy benefits of type-directed synthesis.

This thesis represents the first such extension to the type-directed synthesis system.
My research involved adding product types to the original synthesis framework, which until then
captured only the simply typed lambda calculus with recursion and algebraic datatypes.  In the
process, I heavily revised and expanded both the formal judgments that govern the synthesis
procedure and the code that implements it, integrating additional theorem proving techniques
that pave the way for future language extensions. 

\paragraph{Contributions of this thesis.}
\begin{enumerate}
\item An analysis of the effect of searching only for typed programs or typed programs in
      normal form on the size of the number of programs at a particular type.
\item A revised presentation of the judgments for the original type-directed synthesis
      system.
\item An extension of the original synthesis judgments to include product types and the
      \emph{focusing} technique.
\item Theorems about the properties of the updated synthesis judgments and focusing, including
      proofs of admissibility of focusing and soundness of the full system.
\item Updates to the type-directed synthesis prototype that implement the new judgments.
\item An evaluation of the performance of the updated
      implementation on several canonical programs involving product types.
\item A thorough survey of related work.
\item A discussion of future research avenues for type-directed synthesis with emphasis on
      the observation that examples are refinement types.
\end{enumerate}

\paragraph{Overview.}
The body of this thesis is structured as follows: I begin with
an in-depth look at the original type-directed synthesis system in Section 2.
In Section 3, I introduce the theory underlying synthesis of products and extensively
discuss the focusing technique, which handles product types efficiently.
I describe implementation changes made to add tuples to the synthesis framework and
evaluate the performance of the modified system in Section 4.  In Section 5, I discuss related work,
including extended summaries of other synthesis systems.  Finally,
I outline future research directions in Section 6, with particular emphasis on using intersection and
refinement types that integrate input-output examples into the type system. I conclude in Section 7.
Proofs of theorems presented throughout this thesis appear in Appendix A.

%% file: tds.tex
\section{Type-Directed Synthesis}

\subsection{Overview}
\label{sec:overview}

The following is a brief summary of type-directed synthesis~\cite{tds},
presented loosely within the
\emph{Syntax-Guided Synthesis}~\cite{sygus} framework.

\paragraph{Background theory.}
The system synthesizes over the simply-typed lambda calculus with recursive functions
and non-polymorphic algebraic datatypes.  In practice, it uses
a subset of OCaml with the aforementioned features.  In order to guarantee
that all programs terminate, the language permits only structural recursion.
Pre-defined functions can be made available to the synthesis process if
desired (i.e., providing \emph{map} and \emph{fold} to a synthesis task
involving list manipulation).

\paragraph{Synthesis problem.}
A user specifies the name and type signature of a function to be synthesized.  No
additional structural guidance or ``sketching'' is provided.

\paragraph{Solution specification.}
The function's type information combined with input-output examples
constrain the function to be synthesized. We can treat
this set of examples as a \emph{partial function} that we wish to
generalize into a total function.
Since type-directed synthesis aims to scale to
multi-argument, higher-order functions, these examples can map multiple
inputs, including other functions, to a single output.
Functions are not permissible as output examples, however,
since the synthesis procedure must be able to
decidably test outputs for equality.

\paragraph{Optimality criterion.}
The synthesis problem as currently described lends itself to a very simple
algorithm: create a function that
(1) on an input specified in an example, supplies the corresponding output and (2)
on all other inputs, returns an arbitrary, well-typed value.  To avoid
generating such useless, \emph{over-fitted} programs, type-directed synthesis requires some notion
of a \emph{best} result.  In practice, we create the smallest (measured in AST nodes)
program that satisfies the specification, since a more generic,
recursive solution will lead to a smaller program than one that merely
matches on examples.

\paragraph{Search strategy.}
Type-directed synthesis treats the signature of the
function in the synthesis problem as a theorem to be proved and uses
a modified form of reverse proof search~\cite{atp} that integrates
input-output examples to generate a program.  By the Curry-Howard
isomorphism, a proof of the theorem is a program with the
desired type; if it satisfies the input-output examples, then it
is a solution to the overall synthesis problem.

\paragraph{Style.}
Type-directed synthesis can be characterized as a hybrid algorithm that has
qualities of both deductive and inductive synthesis.
It uses a set of rules to extract the structure of the
input-output examples
into a program, which is reminiscent of deductive synthesis algorithms that use
a similar process on complete specifications.  When guessing variables and
function applications, however, type-directed synthesis generates terms and
checks whether they satisfy the examples, an approach in the style
of inductive strategies like CEGIS~\cite{sketch2}.

\newcommand{\Vt}[1]{\footnotesize \texttt{#1}}
\newcommand{\Vs}[1]{\scriptsize \texttt{#1}}

\begin{figure*}
\lstset{
  basicstyle=\ttfamily,
  breaklines=true
  }
\newcommand{\Ex}[2]{\Vs{{#1}:$\langle${#2}$\rangle$}}
\newcommand{\Jb}{}
\newcommand{\Sb}{\Vt{len : natlist -> nat = ?}}
\newcommand{\Eb}{
\pbox{5cm}{
  \vspace{.5\baselineskip} 
  \Ex{?}{$\cdot$; []~~~~~=> 0} \\
  \Ex{?}{$\cdot$; [3]~~~~=> 1} \\
  \Ex{?}{$\cdot$; [4; 3]~=> 2}
  \vspace{.5\baselineskip} 
}
}
\newcommand{\Db}{\footnotesize (a)~~Initial synthesis problem.}

\newcommand{\Jc}{\footnotesize \textsc{IRefine-Fix}}
\newcommand{\Sc}{\Vt{len (ls : natlist) : nat = ?}}
\newcommand{\Ec}{
\pbox{6cm}{
  \vspace{.5\baselineskip} 
  \Ex{?}{len = ..., ls =~~~~~[]; 0} \\
  \Ex{?}{len = ..., ls =~~~~[3]; 1} \\
  \Ex{?}{len = ..., ls =~[4; 3]; 2}
  \vspace{.5\baselineskip} 
}
}
\newcommand{\Dc}{\footnotesize (b)~~Synthesize a function.}

\newcommand{\Jd}{\footnotesize \textsc{IRefine-Match, EGuess-Ctx}}
\newcommand{\Sd}{
\pbox{8cm}{
  \vspace{.5\baselineskip}
  \Vt{len (ls : natlist) : nat = } \\
  \Vt{\phantom{2em}match ls with} \\
  \Vt{\phantom{2em}| Nil~~~~~~~~~~-> ?$_1$} \\
  \Vt{\phantom{2em}| Cons(hd, tl) -> ?$_2$}
  \vspace{.5\baselineskip}
}
}
\newcommand{\Ed}{
\pbox{9cm}{
  \vspace{.5\baselineskip} 
  \Ex{?$_1$}{len = ..., ls =~~~~~[]~~~~~~~~~~~~~~~~~~; 0} \\
  \Ex{?$_2$}{len = ..., ls =~~~~[3], hd = 3, tl =~~[]; 1} \\
  \Ex{?$_2$}{len = ..., ls = [4; 3], hd = 4, tl = [3]; 2}
  \vspace{.5\baselineskip} 
}
}
\newcommand{\Dd}{\footnotesize (c)~~Synthesize a match statement.}

\newcommand{\Je}{\footnotesize \textsc{IRefine-Ctor}}
\newcommand{\Se}{
\pbox{8cm}{
  \vspace{.5\baselineskip}
  \Vt{len (ls : natlist) : nat = } \\
  \Vt{\phantom{2em}match ls with} \\
  \Vt{\phantom{2em}| Nil~~~~~~~~~~-> O} \\
  \Vt{\phantom{2em}| Cons(hd, tl) -> ?$_2$}
  \vspace{.5\baselineskip}
}
}
\newcommand{\Ee}{
\pbox{9cm}{
  \vspace{.5\baselineskip} 
  \Ex{?$_2$}{len = ..., ls =~~~~[3], hd = 3, tl =~~[]; 1} \\
  \Ex{?$_2$}{len = ..., ls = [4; 3], hd = 4, tl = [3]; 2}
  \vspace{.5\baselineskip} 
}
}
\newcommand{\De}{\footnotesize (d)~~Complete the \Vt{Nil} branch with the \Vt{O} constructor.}

\newcommand{\Jf}{\footnotesize \textsc{IRefine-Ctor}}
\newcommand{\Sf}{
\pbox{8cm}{
  \vspace{.5\baselineskip}
  \Vt{len (ls : natlist) : nat = } \\
  \Vt{\phantom{2em}match ls with} \\
  \Vt{\phantom{2em}| Nil~~~~~~~~~~-> O} \\
  \Vt{\phantom{2em}| Cons(hd, tl) -> S(?$_2$)}
  \vspace{.5\baselineskip}
}
}
\newcommand{\Ef}{
\pbox{9cm}{
  \vspace{.5\baselineskip} 
  \Ex{?$_2$}{len = ..., ls =~~~~[3], hd = 3, tl =~~[]; 0} \\
  \Ex{?$_2$}{len = ..., ls = [4; 3], hd = 4, tl = [3]; 1}
  \vspace{.5\baselineskip} 
}
}
\newcommand{\Df}{\footnotesize (e)~~Synthesize constructor \Vt{S} in the remaining branch.}

\newcommand{\Jg}{\footnotesize \textsc{IRefine-Guess, EGuess-App, EGuess-Ctx}}
\newcommand{\Sg}{
\pbox{8cm}{
  \vspace{.5\baselineskip}
  \Vt{len (ls : natlist) : nat = } \\
  \Vt{\phantom{2em}match ls with} \\
  \Vt{\phantom{2em}| Nil~~~~~~~~~~-> O} \\
  \Vt{\phantom{2em}| Cons(hd, tl) -> S(len~tl)}
  \vspace{.5\baselineskip}
}
}
\newcommand{\Eg}{
\pbox{8cm}{
  \vspace{.5\baselineskip}
  \vspace{.5\baselineskip} 
}
}
\newcommand{\Dg}{\footnotesize (f)~~Synthesize an application to fill the final hole.}

\def\arraystretch{1.5}
\begin{tabular}{|p{1.6cm}|p{5.8cm}|p{7.8cm}|}
\hline
\textbf{\footnotesize Description}         & \Db & \Dc \\ \hline
\textbf{\footnotesize Program}             & \Sb & \Sc \\ \hline
\textbf{\footnotesize Examples}            & \Eb & \Ec \\ \hline
\textbf{\footnotesize Judgment}            & \Jb & \Jc \\ \hline
\end{tabular}

\vspace{\baselineskip}

\begin{tabular}{|p{7.8cm}|p{7.8cm}|}
\hline
\Dd & \De \\ \hline
\Sd & \Se \\ \hline
\Ed & \Ee \\ \hline
\Jd & \Je \\ \hline
\end{tabular}

\vspace{\baselineskip}

\begin{tabular}{|p{7.8cm}|p{7.8cm}|}
\hline
\Df & \Dg \\ \hline
\Sf & \Sg \\ \hline
\Ef & \Eg \\ \hline
\Jf & \Jg \\ \hline
\end{tabular}

\caption{A step-by-step derivation of the list length function in type-directed synthesis.
         A ? character refers to a hole in the program that the synthesis algorithm aims to fill.
         Each example \emph{world}, delimited with $\langle$ and $\rangle$, contains variable
         bindings to the left of the ; and the goal value to the right.  The preceding ?
         indicates the hole to which the example world corresponds. For brevity, we write unary
         numbers in their Arabic equivalents (S (S (O)) is abbreviated as 2) and lists in
         equivalent OCaml syntax (Cons(2, Cons(1, Nil)) is [2; 1]).  The names of all recursive
         functions in scope (in this case \texttt{len}) are always available in the list of variable
         bindings.  They are bound to the partial functions comprising their definitions.  The
         example for \texttt{len} is the initial partial function example at the beginning of
         the synthesis process; it is elided from the example worlds for space.}
\label{fig:example}
\end{figure*}

\subsection{Case Study: List \emph{length}}

Before delving into the technical details of the theory, consider
the process of synthesizing the list \emph{length} function as illustrated in
Figure \ref{fig:example}.

In Figure 1a, we begin with our synthesis problem: a function with its type
signature and a list of input-output examples.  The \texttt{?} preceding each
example represents the hole to which the example corresponds.  Initially,
the entire function to be synthesized comprises a single hole.

Enclosed within angle brackets is each example \emph{world}, which we define
as a pair of (1) bindings of names to values and (2) the \emph{goal} value that should
be taken on by the expression chosen to fill the hole
when the names are bound to those values.  For example,
we can read the expression

\begin{adjustwidth}{0em}{}
\centering
\texttt{?$_1$: $\langle$x = 1, y = 2; 5$\rangle$}
\end{adjustwidth}

\noindent as

\begin{adjustwidth}{1em}{}
\emph{When x = 1 and y = 2, the expression synthesized to fill hole ?$_1$
      should evaluate to 5.}
\end{adjustwidth}

\noindent In our initial synthesis problem in Figure 1a, no names have yet
been bound to values.  We could imagine providing the synthesis instance with
a library of existing functions, like \emph{fold} and \emph{map}, in which case
our example worlds would contain already-bound names at the start of the synthesis
process.

Each goal value is expressed as a partial function mapping input arguments to an
output.  For example,

\begin{adjustwidth}{0em}{}
\centering
\texttt{[1; 2] => inc => [2; 3]}
\end{adjustwidth}

\noindent means that, on inputs \texttt{[1; 2]} and the increment function,
the desired output is \texttt{[2; 3]}.

In Figure 1b, we observe that every example is a partial function mapping
a \texttt{natlist} to a \texttt{nat} and, as such, we synthesize a function
with a \texttt{natlist} argument called \texttt{ls}.  We must update our
examples in kind.  Where before we had an example world of the form

\begin{adjustwidth}{0em}{}
\centering
\texttt{$\langle \cdot$; [4; 3] => 2$\rangle$}
\end{adjustwidth}

\noindent we now extract the value of \texttt{ls} and move it to the list of
names bound to values:

\begin{adjustwidth}{0em}{}
\centering
\texttt{$\langle$ls = [4; 3]; 2$\rangle$}
\end{adjustwidth}

\noindent Since we have now synthesized a function, our hole is of type \texttt{nat} and
the goal value is the remainder of the partial function with \texttt{ls} removed.

In Figure 1c, we synthesize a \texttt{match} statement to break down the structure of
\texttt{ls}.  This creates two holes, one for each branch of the \texttt{match} statement.
We partition our set of examples between the two holes depending on the constructor
of the value bound to \texttt{ls}.  Those examples for which
\texttt{ls = []} constrain the hole for the
\texttt{Nil} branch; all other examples constrain the \texttt{Cons} branch.

In Figure 1d, we turn our attention to the \texttt{Nil} branch.  Since we only have
a single example, it is safe to simply synthesize the value in the example's goal
position (namely the constructor \texttt{O}).
In the other branch, too, every example has the same constructor (\texttt{S}), which
we then generate in our program (Figure 1e).  We update the value in the goal position
accordingly, removing one use of the \texttt{S} constructor.
Finally, with a recursive call to
\texttt{len} on \texttt{tl}, our function is complete.

It is important to note that every function we generate is recursive.  Therefore,
the name of each function, including the
top-level function of the synthesis problem, is available in every example world
for which it is in scope.  Function names are bound to the partial functions
comprising their examples.  Making these names available allows for recursive
function calls.  The examples for \texttt{len} are elided from example worlds in Figure 1
for readability.

\begin{figure}
\centering
\includegraphics[width=.7\linewidth]{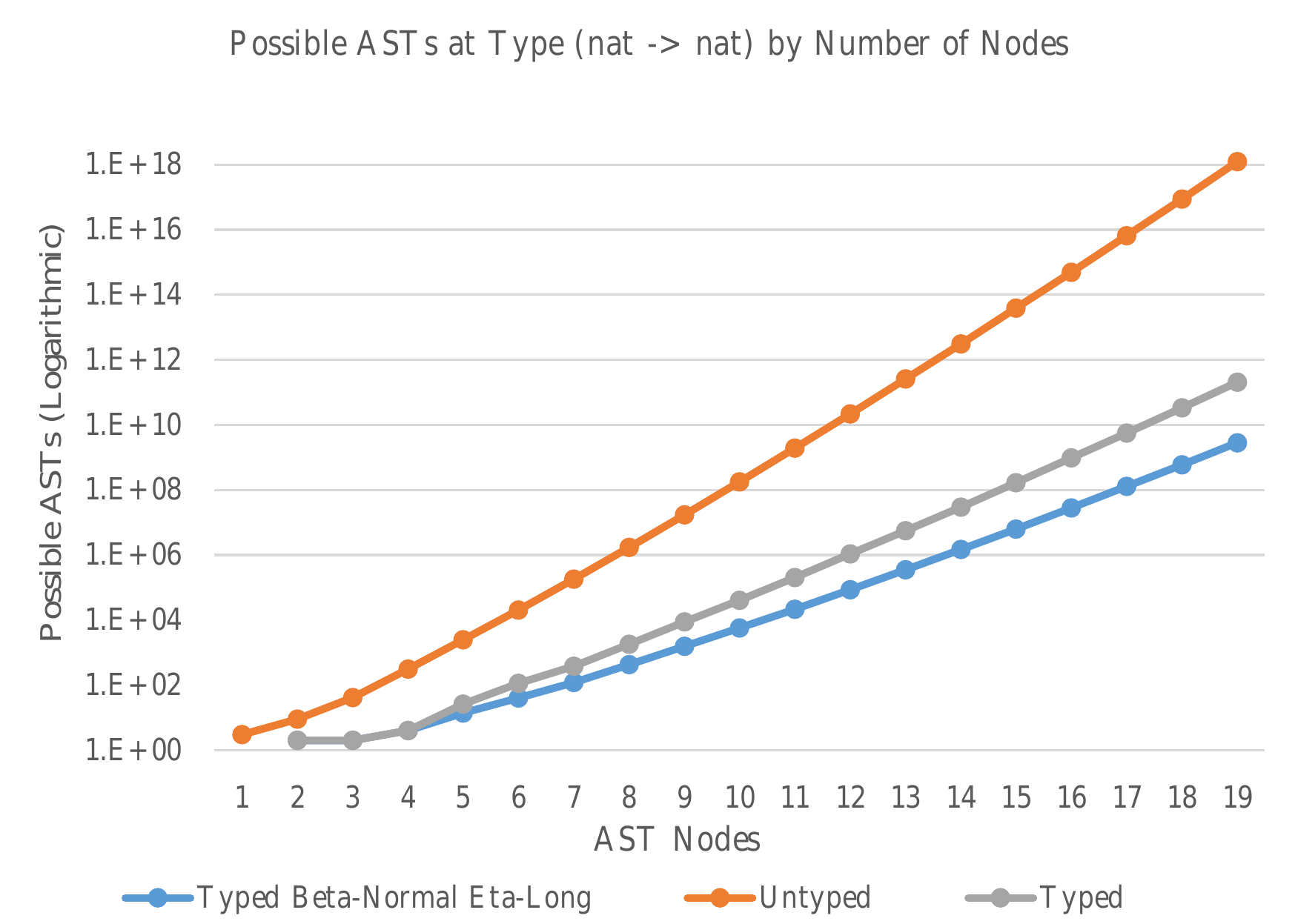}
\caption{The number of possible ASTs (at type nat $\rightarrow$ nat) with a particular number of nodes.
         ASTs were generated over the type-directed synthesis language (the lambda calculus
         with recursive functions and algebraic datatypes).  The top line (orange) includes all
         possible ASTs, while the middle and bottom lines include only typed (grey) and
         typed beta-normal eta-long (blue) ASTs respectively. The scale on the vertical
         axis is logarithmic.}
\end{figure}

\begin{figure*}
\footnotesize
\input{tds-grammar}
\caption{Grammars for the synthesis judgments of type-directed synthesis.}
\end{figure*}

\begin{figure*}
\footnotesize
\input{tds-rules}
\caption{Synthesis judgments for type-directed synthesis.  An expression with a bar
         and an index above it represents a set of expressions sharing the
         same structure.}
\end{figure*}

\subsection{Proof Search}

\paragraph{Search strategy.}

The primary distinguishing quality of type-directed synthesis is its search strategy.
The grammar of expressions $e$ in the background theory is specified in Figure 3. Within
this grammar are myriad ASTs that are not well-typed.  As Figure 2 illustrates, merely
restricting our search to well-typed terms drastically decreases the size of the search space.
Even amongst well-typed terms, there are numerous expressions that are functionally identical.
For example, \EApp[(\EFix[f][x][\DType_1][\DType_1][x])][1] beta-reduces to 1 and
\EFix[f][x][\DType_1][\DType_2][(\EApp[g][x])] eta-reduces to $g$.
If we confine our search to programs in eta-long, beta-normal form, we avoid considering some
of these duplicates and further reduce our search space.
Figure 2 demonstrates the benefit of this further restriction: an additional one to two
orders of magnitude reduction in the size of the search space.

Based on these observations, type-directed syntax avoids searching amongst all expressions
($e$), instead generating well-typed ASTs over a more restrictive grammar: that of 
introduction ($I$) and elimination ($E$) forms.  Doing so guarantees that programs synthesized
are in beta-normal form, since no lambda term can appear on the left side of an application.
If, in addition, we generate only lambdas at function type, programs will also be eta-long.

We can combine these ideas into a proof search algorithm that begins with a
synthesis problem as in Figure 1 and uses type information, along with the
structure of examples, to generate programs.  The synthesis judgments for this
algorithm appear in Figure 4.

\paragraph{Refining introduction forms.}

The judgment for producing introduction forms, which is also the top-level
synthesis judgment, is of the form

\begin{adjustwidth}{0em}{}
\centering
$\RefineO$
\end{adjustwidth}

\noindent which states:

\begin{adjustwidth}{1em}{}
\emph{Given constructors \CtorCtx~and names bound to types \TypeCtx, we synthesize
      introduction form \Intro~at type \DType~conforming to the examples in worlds \WorldCtx.}
\end{adjustwidth}

The \textsc{IRefine-Fix} rule extracts the structure of a partial function example
as in Figure 1b, synthesizing a fixpoint and creating a new subproblem for the fixpoint's
body that can refer to the argument and make recursive calls.
The \textsc{IRefine-Ctor} rule observes that every
example shares the same constructor and therefore synthesizes the constructor in the program,
likewise creating subproblems for each of the constructor's arguments.

The \textsc{IRefine-Match} rule generates a \texttt{match} statement, guessing an elimination-form
scrutinee at base type on which to pattern-match.
The rule creates a sub-problem for each branch of the match statement -
that is, one for every constructor of the scrutinee's type.
The existing set of examples is partitioned among these sub-problems according to the value that
the scrutinee takes on when evaluated in each example world.  Since the values of the bound
variables may vary from world to world, the scrutinee can evaluate to different constructors
in different contexts.  This evaluation step determines which examples correspond to which
branches, allowing us to constrain the sub-problems and continue with the synthesis process.

Finally, at base type (\TBase), we may also guess an elimination form, captured by
the \textsc{IRefine-Guess} rule.  The restriction that this only occurs at base type
helps enforce the property that we only generate eta-long programs.

\paragraph{Guessing elimination forms.}
The judgment form for guessing elimination forms

\begin{adjustwidth}{0em}{}
\centering
$\EGuessO[\CtorCtx][\TypeCtx][\DType][\Elim]$
\end{adjustwidth}

\noindent states:

\begin{adjustwidth}{1em}{}
\emph{Given constructors \CtorCtx~and names bound to types \TypeCtx, we synthesize
      elimination form \Elim~at type \DType.}
\end{adjustwidth}

\noindent Observe that this judgment form does not include examples.  Unlike the introduction
forms, we cannot use the structural content of the examples to inform
the elimination forms we guess.  Instead, we may only ensure that the elimination
form that we generate obeys the examples.  This requirement appears in the additional condition
of the \textsc{IRefine-Guess} rule

\begin{adjustwidth}{0em}{}
\centering
$\forall i \in n,~ \ExamCtx_i(\Elim) \SssStar \Exam_i$
\end{adjustwidth}

\noindent reading:

\begin{adjustwidth}{1em}{}
\emph{For each of our $n$ examples, the elimination form \Elim~must, when substituted with the
      value bindings in $\ExamCtx_i$, evaluate to the corresponding goal value $\DExam_i$.}
\end{adjustwidth}

The \textsc{EGuess-App} rule guesses a function application by recursively guessing
a function and an argument.  The function must be an elimination form, which ensures that
we only guess beta-normal programs.  The argument may be an introduction form, but the
call back into an \textsc{IRefine} rule does so without examples.  Finally,
the base case \textsc{EGuess-Ctx} guesses a name from the context \TypeCtx.

\paragraph{Nondeterminism.}
These rules are highly nondeterministic: at any point, several rules apply.
This is especially pertinent at base type, where we could refine a constructor or
match statement while guessing an elimination form.
Figure 1 traces only one of many paths through the expansive
search space that this nondeterminism creates.

Nondeterminism provides the
flexibility for an optimality condition (Section \ref{sec:overview}) to select the
best possible program.  Although we could continually apply \textsc{IRefine} rules
until we fully extract the structure of the examples, doing so would generate a
larger than necessary program that has been \emph{overfitted} to the examples.  Instead,
the \textsc{EGuess} rules allow us to make recursive calls that create smaller,
more general programs.

\paragraph{Relationship with other systems.}
The synthesis judgments are closely related to two other systems: the sequent calculus
and bidirectional type-checking.  We can reframe the judgment

\begin{adjustwidth}{1em}{}
\centering
$\RefineO[\CtorCtx][\HasType[\DVar_1][\DType_1], ..., \HasType[\DVar_n][\DType_n]][\WorldCtx][\DType]$
\end{adjustwidth}

\noindent as the sequent

\begin{adjustwidth}{1em}{}
\centering
$\DType_1, ..., \DType_n \Longrightarrow \DType$
\end{adjustwidth}

\noindent if we elide the proof terms ($\DVar_1$ through $\DVar_n$ and $\Intro$) and examples.
Many of our rules (particularly \textsc{IRefine-Fix} and \textsc{EGuess-App})
directly mirror their counterparts in Gentzen's characterization of the
sequent calculus~\cite{gentzen}.  The only modifications are the addition of proof terms
and bookkeeping to manage examples.  The rules that
handle algebraic datatypes and pattern matching (\textsc{IRefine-Ctor} and \textsc{IRefine-Match})
loosely resemble logical disjunction, which Gentzen also describes.

The type-directed synthesis system also resembles bidirectional typechecking.  Where
traditional bidirectional typechecking rules are functions that synthesize or check a type given
a context and an expression, however, our judgments synthesize an expression
from a context and a type.  The \textsc{IRefine} judgments correspond to rules where a type
would be checked and the \textsc{EGuess} judgments to rules where a type would be synthesized.

%% file: tds-grammar.tex

\DeclareDocumentCommand{\EOMatch}{O{\DExp}
                                  O{\Overbar{i \in l}{\DCtor_i~\ETupleRange{\DVar}\rightarrow e_i}}}
                       {\ensuremath{\mathsf{match}~{#1}~\mathsf{with}~{#2}}}

\begin{tabular}{l l l}

\DType &
\GEq \TBase \GBar \TFunc{\DType_1}{\DType_2} &
(Types: base types and function types)\\

\DVal &
\parbox[t]{5cm}{
\GEq \ECtor[\DCtor][\ETupleRange{\DVal}] \\
\phantom{1em}\GBar \EFix \GBar \Partial
}
&
(Values: constructors, functions, and partial functions) \\

\Partial &
\GEq \Overbar{i \in m}{\DVal_i \Rightarrow \DExam_i} &
(Partial functions: a finite map from inputs to outputs)\\

\DExam &
\GEq \ECtor[\DCtor][\ETupleRange{\DExam}] \GBar \Partial &
(Examples)\\

\DExp &
\parbox[t]{6cm}{
\GEq \DVar \GBar \ECtor[\DCtor][\ETupleRange{\DExp}] \GBar \Partial \\
\phantom{1em}\GBar \EFix \GBar \EApp \\
\phantom{1em}\GBar \EOMatch
}
&
(Expressions)\\

& & \\

\Elim &
\GEq \DVar \GBar \EApp[\Elim][\Intro] &
(Elimination forms)\\

\Intro &
\parbox[t]{6cm}{
\GEq \Elim \GBar \ECtor[\DCtor][\ETupleRange{\Intro}] \\
\phantom{1em}\GBar \EFix[f][x][\DType_1][\DType_2][\Intro] \\
\phantom{1em}\GBar \EOMatch[\Elim][\Overbar{i \in l}{\DCtor_i~\ETupleRange{\DVar} \rightarrow \Intro_i}]
}
 &
(Introduction forms)\\

& & \\

\CtorCtx &
\GEq \GEmp \GBar \CtorCtx, \HasType[\DCtor][\CtorType{\DType_1 \TStar ... \TStar \DType_n}] &
(Constructor context: a map from constructor names to types) \\

\TypeCtx &
\GEq \GEmp \GBar \TypeCtx, \HasType[\DVar][\DType] &
(Type context: bindings of names to types) \\

\ExamCtx &
\GEq \GEmp \GBar \ExamCtx, \Refines[\DVar][\DExam] &
(Example context: names refined by examples) \\

\WorldOne &
\GEq \World &
(A single world: an example context and result example) \\

\WorldCtx &
\GEq \GEmp \GBar \WorldCtx, \WorldOne &
(A set of worlds) \\

\end{tabular}

%% file: tds-rules.tex

\begin{gather*}
\boxed{\RefineO} \\ \\
\inferrule[IRefine-Guess]{
    \EGuessO[\CtorCtx][\TypeCtx][\TBase][\Elim] \\
    \forall i \in n,~ \ExamCtx_i(\Elim) \SssStar \Exam_i
}{
    \RefineO[\CtorCtx][\TypeCtx][\Overbar{i \in n}{\World[\ExamCtx_i][\Exam_i]}][\TBase][\Elim]
}~~~~~~~~~~~~~~
\inferrule[IRefine-Ctor]{
    \HasType[\DCtor][\TFunc{\DType_1 \TStar ... \TStar \DType_m}{\TBase}] \in \CtorCtx \\
    \forall j \in m, \RefineO[\CtorCtx][\TypeCtx
          ][\Overbar{i \in n}{\World[\ExamCtx_i][\Exam_{(i, j)}]}][\DType_j][\Intro_j]
}{
    \RefineO[\CtorCtx][\TypeCtx
          ][\Overbar{i \in n}{\World[\ExamCtx_i][\ECtor[\DCtor][(\Exam_{(i, 1)}, ..., \Intro_{(i, m)})
         ]]}][\TBase][\ECtor[\DCtor][\ETupleRange{\Intro}]]
} \\ \\
\inferrule[IRefine-Fix]{
    \RefineO[\CtorCtx
          ][\TypeCtx, \HasType[\DFun][\TFunc{\DType_1}{\DType_2}], \HasType[\DVar][\DType_1]
          ][\Overbar{(i, k) \in (n, m_i)}{\World[\ExamCtx_i,
                                             \Refines[f][\Overbar{j \in m_i}{\DVal_{(i, j)}
                                                         \Rightarrow \DExam_{(i, j)}}],
                                             \Refines[x][\DVal_{(i, k)}]
                                           ][\DExam_{(i, k)}]}
          ][\DType_2
          ][\Intro]
}{
    \RefineO[\CtorCtx
          ][\TypeCtx
          ][\Overbar{i \in n}{\World[\ExamCtx_i
                                   ][\Overbar{j \in m_i}{\DVal_{(i, j)} \Rightarrow \DExam_{(i, j)}}]}
          ][\TFunc{\DType_1}{\DType_2}
          ][\EFix[f][x][\DType_1][\DType_2][\Intro]]
} \\ \\
\inferrule[IRefine-Match]{
    \EGuessO[\CtorCtx][\TypeCtx][\TBase][\Elim] \\
    \forall j \in m, ~\HasType[\DCtor_j][\TFunc{\DType_{(j, 1)} \TStar ... \TStar \DType_{(j, l)}}{\TBase}] \in \CtorCtx \\
    \forall j \in m, ~\RefineO[\CtorCtx][\TypeCtx, \HasType[\DVar][\DType_{(j, 1)}], ...,
                                                   \HasType[\DVar][\DType_{(j, l)}]
                             ][\{\World[\ExamCtx_i, \Refines[\DVar_1][\DVal_1], ...,
                                                    \Refines[\DVar_l][\DVal_l]
                                       ][\Exam_i]
                               ~\big|~
                               \ExamCtx_i(\Elim) \SssStar \DCtor_j~(\DVal_1, ..., \DVal_l)\}
                             ][\DType][\Intro_j]
}{
    \RefineO[\CtorCtx][\TypeCtx
          ][\Overbar{i \in n}{\World[\ExamCtx_i][\Exam_i]}][\DType
          ][\EMatch[\Elim][\Overbar{j \in m}{\DCtor_j~(\DVar_1, ..., \DVar_l) \rightarrow \Intro_j}]]
}
\end{gather*}

\begin{gather*}
\boxed{\EGuessO[\CtorCtx][\TypeCtx][\DType][\Elim]}\\ \\
\inferrule[EGuess-Ctx]{
    \HasType[\DVar][\DType] \in \TypeCtx
}{
    \EGuessO[\CtorCtx][\TypeCtx][\DType][\DVar]
}~~~~~~~~~~~~~~~~~~~~
\inferrule[EGuess-App]{
    \EGuessO[\CtorCtx][\TypeCtx][\TFunc{\DType_1}{\DType}][\Elim] \\
    \RefineO[\CtorCtx][\TypeCtx][\cdot][\DType_1][\Intro]
}{
    \EGuessO[\CtorCtx][\TypeCtx][\DType][\EApp[\Elim][\Intro]]
}
\end{gather*}

%% file: theory.tex
\section{Synthesis of Products}

\subsection{Overview of Products}

In this section, I describe the process of adding k-ary (where k $>$ 1) tuples
to the type-directed synthesis framework.  The syntax of tuples is identical to
that of OCaml.  A tuple of k values is written:

\begin{adjustwidth}{1em}{}
\centering
$(\DVal_1, ..., \DVal_k)$
\end{adjustwidth}

\noindent Tuples are eliminated using projection, which, in our rules, follows syntax
similar to that of Standard ML.  The j$^{th}$ projection of a k-ary tuple (where $1 \leq j \leq k$)

\begin{adjustwidth}{0em}{}
\centering
$\EProj[j][(\DVal_1, ..., \DVal_k)]$
\end{adjustwidth}

\noindent evaluates to

\begin{adjustwidth}{0em}{}
\centering
$\DVal_j$
\end{adjustwidth}

As in OCaml, a
tuple's type is written as the ordered cartesian product of the types of its constituents.
For example, the type

\begin{adjustwidth}{0em}{}
\centering
\TTuple{\mathsf{nat} \TStar \mathsf{list} \TStar \mathsf{bool}}
\end{adjustwidth}

\noindent denotes the type of a tuple whose members are, in order, a natural number, a list, and
a boolean.

\subsection{Tuples as a Derived Form}

It is entirely possible to encode tuples as a derived form in the original
type-directed synthesis framework.  We could declare an algebraic datatype with a single
variant that stores k items of the necessary types.  Projection would entail matching on
a ``tuple'' to extract its contents.  Although this strategy will successfully
integrate a functional equivalent of products into the system, first class tuples
are more desirable for several reasons.

\paragraph{Convenience.} Since the existing system lacks polymorphism,
we would need to declare a new algebraic type for every
set of types we wish to combine into a product.  Even with polymorphism, we would still need a new
type for each arity we might desire.  This approach would swiftly become
tedious to a programmer and, worse, would lead the synthesizer to write incomprehensible
programs filled with computer-generated type declarations and corresponding
match statements.

\paragraph{Orthogonality.}
Implementing tuples as a derived form is an inelegant and syntactically-heavy way to
add an essential language feature to the system.
In contrast, first-class tuples are orthogonal,
meaning we can add, remove, or modify other features without affecting products.

\paragraph{Cleaner theory.}
First class tuples simplify a number of other language features, including
constructors and match statements.  In the original type-directed synthesis
system, variable-arity constructors were necessary to implement fundamental
recursive types like lists.  First class tuples can replace this functionality,
streamlining the theory behind the entire system.

\paragraph{Efficient synthesis.}
Theorem-proving machinery can take advantage of the
logical properties of product types to produce a more efficient
synthesis algorithm (see Section \ref{sec:focusing}).  Although we could
specially apply these techniques to single-variant, non-recursive
algebraic datatypes (the equivalent of tuples), we can simplify
the overall synthesis algorithm by delineating a distinct product type.

\paragraph{Informative matching.}
The original synthesis implementation only generates a match statement if
doing so partitions the examples over two or more branches, a restriction
known as \emph{informative matching}. The rationale
for this requirement is that pattern matching is only useful if it
separates our synthesis task into smaller subproblems, each with fewer examples.  A
single-constructor variant would never satisfy this constraint, meaning that
we would need another special case in order to use match statements to
``project'' on our tuple derived form.

\begin{figure*}
\footnotesize
\input{products-grammar}
\caption{Updated grammars for type-directed synthesis with products.  Changes are highlighted.}
\label{fig:productsgrammar}
\end{figure*}

\begin{figure*}
\footnotesize
\input{focusing-rules}
\caption{Focusing rules for type-directed synthesis with products.}
\label{fig:focusingrules}
\end{figure*}

\begin{figure*}
\footnotesize
\input{products-rules}
\caption{The updated judgments for type-directed synthesis with products.  Altered judgments
         have been highlighted.}
\label{fig:productsrules}
\end{figure*}

\subsection{Focusing}
\label{sec:focusing}

\paragraph{Theorem proving strategy.}
Product types are the equivalents of logical conjunction, which itself is
\emph{left-invertible}~\cite{atp}.  That is, when we know that
$A \land B$ is true, we can eagerly conclude that
$A$ and $B$ are individually true without loss of information.  Should we
later need the fact that $A \land B$ is true, we can easily re-prove
it.

We can turn this invertibility into an efficient strategy for theorem-proving called
\emph{focusing}.  When we add a value of product type to the context (our proof-term
equivalent of statements we know to be true) we can eagerly project on it in order
to break it into its non-product type constituents.  Since products are invertible,
we can always reconstruct the original entity from its projected components if the need arises.
Where many parts of the type-directed synthesis strategy require us to guess derivations
that might not lead to a candidate program,
focusing allows us to advance our proof search in a way that will never
need to be backtracked.

\paragraph{Synthesis strategy.}
Integrating focusing into type-directed synthesis requires several fundamental changes
to the existing theory, including the addition of an entirely new focusing judgment
form (Figure \ref{fig:focusingrules}) and additional ``focusing'' contexts in the \textsc{IRefine} and
\textsc{EGuess} judgments (Figure \ref{fig:productsrules}).  The grammar of the type context \TypeCtx~has
subtly changed: rather than binding names to types, it now binds elimination forms \Elim~(Figure \ref{fig:productsgrammar}).
This alteration allows the context to store, not only a value
$(\HasType[\DVar][\DType_1 \TStar \DType_2])$, but also the results of focusing:
$(\HasType[\EProj[1][\DVar]][\DType_1])$ and $(\HasType[\EProj[2][\DVar]][\DType_2])$.

Two new contexts share the same structure as \TypeCtx: a \emph{focusing context} \FocusCtx~
and an \emph{auxiliary context} \AuxCtx.  We use the contexts as follows: when a new name or,
more generally, elimination form is to be added to the context, we first insert it into
\FocusCtx.  Items in \FocusCtx~are repeatedly focused and reinserted into \FocusCtx~until
they cannot be focused any further, at which point they move into \TypeCtx.  The terms
that we project upon in the course of focusing are no longer useful for the synthesis process but are
critical for proving many properties about our context; they are therefore permanently moved into
\AuxCtx.

\paragraph{Invariants.}
We can crystalize this workflow into several invariants about the contexts:
\begin{enumerate}
\item New names and elimination forms are added to \FocusCtx.
\item Elimination forms in \TypeCtx~cannot be focused further.
\item No other rule can be applied until focusing is done.
\item Only elimination forms in \TypeCtx~can be used to synthesize expressions.
\end{enumerate}

\noindent We have already discussed the justification for the first two invariants.  
The third invariant ensures that we prioritize backtracking-free
focusing over the nondeterministic choices inherent in applying other rules.  By
always focusing first, we add determinism to our algorithm
and advance our proof search as far as possible before making
any decisions that we might later have to backtrack.  To enforce this invariant,
all of the \textsc{IRefine} and \textsc{EGuess} rules that synthesize expressions
require \FocusCtx~to be empty.

The final invariant ensures
that we never generate terms from \AuxCtx~or \FocusCtx~that may not have been entirely
focused, preserving the property that we only synthesize normal forms.  Otherwise, we
could synthesize both a term $(\HasType[\DVar][\DType_1 \TStar \DType_2])$ and the
eta-equivalent expression $(\HasType[(\EProj[1][\DVar], \EProj[2][\DVar])][\DType_1 \TStar \DType_2])$.
This invariant implies an eta-long normal form for tuples: among the aforementioned
two expressions,
our algorithm will always synthesize the latter.

\paragraph{Focusing judgments.}
Special \textsc{IRefine-Focus} and \textsc{EGuess-Focus} rules allow the focusing
process to advance to completion so that synthesis can continue.  They make calls to a
separate focusing judgment (Figure \ref{fig:focusingrules}) that operates on contexts and example worlds.
The judgment

\begin{adjustwidth}{1em}{}
\centering
\IFocus[
    \RefineCtx[\CtorCtx_1][\TypeCtx_1][\AuxCtx_1][\FocusCtx_1][\WorldCtx_1]
  ][\RefineCtx[\CtorCtx_2][\TypeCtx_2][\AuxCtx_2][\FocusCtx_2][\WorldCtx_2]
  ]
\end{adjustwidth}

\noindent states that

\begin{adjustwidth}{1em}{}
\emph{The contexts $\CtorCtx_1$, $\TypeCtx_1$, $\AuxCtx_1$, and $\FocusCtx_1$~along with
      example worlds $\WorldCtx_1$~can be focused into the contexts
      $\CtorCtx_2$, $\TypeCtx_2$, $\AuxCtx_2$, and $\FocusCtx_2$~along with
      example worlds $\WorldCtx_2$.}
\end{adjustwidth}

\noindent The focusing rules move any elimination form of non-product type from \FocusCtx~to
\TypeCtx~without modifying any other contexts.  On product types, the rule \textsc{Focus-Tuple}
projects on an elimination form in \FocusCtx, returning the resulting expressions to \FocusCtx~
for possible additional focusing.  It splits the examples of the focused expression amongst
the newly-created projections and moves the now-unnecessary original expression into \AuxCtx.

\subsection{Changes to Synthesis Judgments}

The updated grammars, focusing rules, and judgments for type-directed synthesis with
products appear in Figures
\ref{fig:productsgrammar}, \ref{fig:focusingrules}, and \ref{fig:productsrules}
respectively.  Below are summaries
of the changes that were made to the original theory in the process of
adding products.

\paragraph{Tuple and projection expressions.}

Expressions for tuples and projection have been added to the grammar of
expressions (\DExp).  Likewise, the product type has been added to the
grammar of types (\DType).  Tuple creation is an introduction form with
projection as the corresponding elimination form.

\paragraph{Single-argument constructors.}
Now that tuples exist as a separate syntactic construct, the variable-arity constructors
of the original type-directed synthesis judgments are redundant: creating a constructor
with an argument of product type achieves the same effect.  Without products, variable-arity
constructors were essential for defining recursive types that stored values (i.e., lists),
but this functionality
can now entirely be implemented using tuples. Therefore, in the updated
judgments, all constructors have a single argument, enormously simplifying the
\textsc{IRefine-Ctor} and \textsc{IRefine-Match} rules.

One drawback of this choice is that pattern matching can no longer deconstruct
both a constructor and the implicit tuple embedded inside it as in the old judgments.
Instead, pattern matching always produces a single variable that must be focused
and projected upon in order to extract its components.
Additional pattern-matching machinery could be added
to the theory (and is present in the implementation) to resolve this
shortcoming.

\paragraph{Unit type.}
Now that all constructors require a single argument, what is the base case in an
inductive datastructure?  Previously, constructors with no arguments (\texttt{O}, \texttt{Nil}, etc.)
served this purpose, but the new rules require them to have an argument as well.  To
resolve this problem, a unit type (\TUnit) and unit value (written \EUnit) have been introduced.
Former no-argument constructors now have type \TUnit.  A straightforward
\textsc{IRefine-Unit} rule synthesizes the unit value at unit type.

\paragraph{Tuple and projection judgments.}
A new \textsc{IRefine-Tuple} judgment synthesizes a tuple at product type
when all examples are tuple values with the same arity, inheriting much of the behavior previously
included in the \textsc{IRefine-Ctor} rule.  It creates one subproblem for every
value contained within the tuple and partitions the examples accordingly.
Notably, there is no corresponding \textsc{EGuess-Proj} judgment.  All projection
instead occurs during focusing.

\paragraph{Corrected application judgment.}
The absence of an \textsc{EGuess-Proj} rule required changes to the
\textsc{EGuess-App} judgment.  As a motivating example, consider the
list \emph{unzip} function:

\begin{adjustwidth}{1em}{}
  \small\texttt{let unzip (ls:pairlist) : list * list =} \\
  \small\texttt{\phantom{1em}match ls with} \\
  \small\texttt{\phantom{1em}| [] -> ([], [])} \\
  \small\texttt{\phantom{1em}| (a, b) :: tl -> (a :: \hl{\#1 (unzip tl)}, b :: \hl{\#2 (unzip tl)})}
\end{adjustwidth}

\noindent Synthesizing \emph{unzip} requires projecting on a function application
(highlighted above).

The application rule in the original type-directed synthesis judgments
is reproduced below. It has been modified slightly to include the new context structure
accompanying the updated judgments in Figure \ref{fig:productsrules}:

\begin{adjustwidth}{0cm}{}
\centering
\ensuremath{
\footnotesize
\inferrule{
    \EGuess[\CtorCtx][\TypeCtx][\AuxCtx][\cdot][\TFunc{\DType_1}{\DType}][\Elim] \\
    \Refine[\CtorCtx][\TypeCtx][\AuxCtx][\cdot][\cdot][\DType_1][\Intro]
}{
    \EGuess[\CtorCtx][\TypeCtx][\AuxCtx][\cdot][\DType][\EApp[\Elim][\Intro]]
}
}
\end{adjustwidth}

\noindent This judgment requires that we synthesize applications at the goal type
and immediately use them to solve the synthesis problem.  In \emph{unzip}, however,
we need to synthesize an application at a product type and subsequently project
on it, meaning that, with this version of the \textsc{EGuess-App} judgment, we
cannot synthesize \emph{unzip}.

Alternatively, we could add a straightforward \textsc{EGuess-Proj} judgment
that would allow us to project on applications:

\begin{adjustwidth}{0cm}{}
\centering
\ensuremath{
\footnotesize
\inferrule{
   \EGuess[\CtorCtx][\TypeCtx][\AuxCtx][\cdot][\TTuple{\DType_1 \TStar ... \TStar \DType_m}][\Elim] \\
   1 \leq k \leq m
}{
    \EGuess[\CtorCtx][\TypeCtx][\AuxCtx][\cdot][\DType_k][\EProj[k][\Elim]]
}
}
\end{adjustwidth}

\noindent Doing so, however, undermines the invariant that we only generate normal
forms.  Suppose we wished to synthesize an expression at type $\DType_1 \TStar \DType_2$.
We could either synthesize an application
$\HasType[\EApp[\Elim][\Intro]][\DType_1 \TStar \DType_2]$
or the corresponding eta-long version
$\HasType[(\EProj[1][(\EApp[\Elim][\Intro])],
            \EProj[2][(\EApp[\Elim][\Intro])])][\DType_1 \TStar \DType_2]$.

For a solution to this quandary, we need to return to the underlying proof theory.  In the
sequent calculus, the left rule for implication, which corresponds to our
\textsc{EGuess-App} judgment, appears as below~\cite{gentzen,atp}:

\begin{adjustwidth}{0cm}{}
\centering
\ensuremath{
\footnotesize
\inferrule{
   (1)~\Gamma, A \rightarrow B \Longrightarrow A \\
   (2)~\Gamma, A \rightarrow B, B \Longrightarrow C
}{
   (3)~\Gamma, A \rightarrow B \Longrightarrow C
}
}
\end{adjustwidth}

\noindent That is:

\begin{adjustwidth}{1em}{}
\emph{If the context contains proof that A implies B and (1) we can prove A and
      (2) if the context also contains a proof of B, then we can prove C, then (3)
      we can prove C.}
\end{adjustwidth}

\noindent This rule suggests a slightly different application rule, namely the one
in Figure 7:

\begin{adjustwidth}{0cm}{}
\centering
\ensuremath{\footnotesize \inferrule[EGuess-App]{
    \Refine[\CtorCtx][\TypeCtx, \HasType[\Elim_1][\TFunc{\DType_1}{\DType_2}]
          ][\AuxCtx][\cdot][\cdot][\DType_1][\Intro_1] \\
    \EGuess[\CtorCtx][\TypeCtx, \HasType[\Elim_1][\TFunc{\DType_1}{\DType_2}]
          ][\AuxCtx][\HasType[\EApp[\Elim_1][\Intro_1]][\DType_2]][\DType][\Elim]
}{
    \EGuess[\CtorCtx][\TypeCtx, \HasType[\Elim_1][\TFunc{\DType_1}{\DType_2}]
          ][\AuxCtx][\cdot][\DType][\Elim]
}
}
\end{adjustwidth}

\noindent The new \textsc{EGuess-App} rule allows us to generate an application at any type,
not just the goal type, and to add it to the context.  We then focus the application,
thereby projecting on it as necessary.  This application now becomes available for use
in the synthesis problem on which we are currently working.  In effect, we take a forward
step in the context by guessing an application.

Note that, although the new application judgment corresponds more closely to the proof theory,
the application judgment in the original type-directed synthesis rules was not incorrect.  Without
projection, the only possible use for a function application in the context would be to
immediately use it to satisfy the synthesis goal.  The application judgment therefore
simply short-circuited the process of first adding an application to the context and then using
it via \textsc{EGuess-Ctx}.

\begin{figure*}

\begin{mathpar}
\footnotesize
\inferrule[Focus-Closure-Base]{
}{
\IFocusStar[
    \RefineCtx[\CtorCtx_1][\TypeCtx_1][\AuxCtx_1][\FocusCtx_1][\WorldCtx_1]
  ][\RefineCtx[\CtorCtx_1][\TypeCtx_1][\AuxCtx_1][\FocusCtx_1][\WorldCtx_1]
  ]
}

\inferrule[Focus-Closure-Step]{
\IFocus[
    \RefineCtx[\CtorCtx_1][\TypeCtx_1][\AuxCtx_1][\FocusCtx_1][\WorldCtx_1]
  ][\RefineCtx[\CtorCtx_2][\TypeCtx_2][\AuxCtx_2][\FocusCtx_2][\WorldCtx_2]
  ] \\
\IFocusStar[
    \RefineCtx[\CtorCtx_2][\TypeCtx_2][\AuxCtx_2][\FocusCtx_2][\WorldCtx_2]
  ][\RefineCtx[\CtorCtx_3][\TypeCtx_3][\AuxCtx_3][\FocusCtx_3][\WorldCtx_3]
  ]
}{
\IFocusStar[
    \RefineCtx[\CtorCtx_1][\TypeCtx_1][\AuxCtx_1][\FocusCtx_1][\WorldCtx_1]
  ][\RefineCtx[\CtorCtx_3][\TypeCtx_3][\AuxCtx_3][\FocusCtx_3][\WorldCtx_3]
  ]
}

\end{mathpar}

\caption{Judgments for the transitive closure of focusing.}
\label{fig:focusclosure}
\end{figure*}

\begin{figure*}
\footnotesize
\begin{mathpar}
\inferrule[T-Var]{
    \HasType[x][\DType] \in \TypeCtx
}{
    \TypeCtx \vdash \HasType[x][\DType]
}

\inferrule[T-Abs]{
    \TypeCtx, \HasType[f][\DTFunc], \HasType[x][\DType_1] \vdash \HasType[\DExp][\DType_2]
}{
    \TypeCtx \vdash \HasType[\EFix[f][x][\DType_1][\DType_2][\DExp]][\DTFunc]
}

\inferrule[T-App]{
    \TypeCtx \vdash \HasType[\DExp_1][\DTFunc] \\
    \TypeCtx \vdash \HasType[\DExp_2][\DType_1]
}{
    \TypeCtx \vdash \HasType[\EApp[\DExp_1][\DExp_2]][\DType_2]
}

\\

\inferrule[T-Unit]{
}{
    \TypeCtx \vdash \HasType[\EUnit][\TUnit]
}

\inferrule[T-Tuple]{
    \forall~i \in m,~\TypeCtx \vdash \HasType[\DExp_i][\DType_i]
}{
    \TypeCtx \vdash \HasType[\ETupleRange{\DExp}][\TTupleRange]
}

\inferrule[T-Proj]{
    \TypeCtx \vdash \HasType[\DExp][\TTupleRange] \\ 1 \leq k \leq m
}{
    \TypeCtx \vdash \HasType[\EProj[k][\DExp]][\DType_k]
}

\\

\inferrule[T-Ctor]{
    \HasType[\DCtor][\TFunc{\DType}{\TBase}] \in \CtorCtx \\
    \TypeCtx \vdash \HasType[\DExp][\DType]
}{
    \TypeCtx \vdash \HasType[\EApp[\DCtor][\DExp]][\TBase]
}

\inferrule[T-Match]{
    \forall~i \in m,~\HasType[\DCtor_i][\TFunc{\DType_i}{\TBase}] \in \CtorCtx \\
    \TypeCtx \vdash \HasType[\DExp][\TBase] \\
    \forall~i \in m,~\TypeCtx, \HasType[x][\DType_i] \vdash \HasType[\DExp_i][\DType]
}{
    \TypeCtx \vdash \HasType[\EMatch][\DType]
}
\end{mathpar}
\caption{Typing judgments for the simply typed lambda calculus with recursion,
         algebraic datatypes, and products.}
\label{fig:types}
\end{figure*}

\begin{figure*}

\begin{mathpar}
\footnotesize
\inferrule[S-Ctor]{
    \DExp \Sss \DExp'
}{
    \ECtor[\DCtor][\DExp] \Sss \ECtor[\DCtor][\DExp']
}

\inferrule[S-Tuple]{
    \DExp_{k} \Sss \DExp_{k}'
}{
    (\DVal_1, ..., \DExp_{k}, ..., \DExp_n) \Sss (\DVal_1, ..., \DExp_k', ..., \DExp_n)
}

\inferrule[S-Proj1]{
    \DExp \Sss \DExp'
}{
    \EProj[k][\DExp] \Sss \EProj[k][\DExp']
}

\inferrule[S-Proj2]{
    1 \leq k \leq m
}{
    \EProj[k][\ETupleRange{\DExp}] \Sss \DExp_k
}

\\

\inferrule[S-App1]{
    \DExp_1 \Sss \DExp_1'
}{
    \EApp[\DExp_1][\DExp_2] \Sss \EApp[\DExp_1'][\DExp_2]
}

\inferrule[S-App2]{
}{
    \EApp[(\EFix[f][x][\DType_1][\DType_2][\DExp_1])][\DExp_2] \Sss
    \DExp_1\left[x \mapsto \DExp_2, f \mapsto \EFix[f][x][\DType_1][\DType_2][\DExp_1] \right]
}

\\

\inferrule[S-Match1]{
    \DExp \Sss \DExp'
}{
    \EMatch \Sss \EMatch[\DExp']
}

\inferrule[S-Match2]{
    1 \leq k \leq m
}{
    \EMatch[\ECtor[\DCtor_k][\DVal]] \Sss
    \DExp_k \left[x \mapsto \DVal\right]
}

\end{mathpar}
\caption{Small-step semantics for the simply typed lambda calculus with recursion,
         algebraic datatypes, and products.}
\label{fig:smallstep}
\end{figure*}

\begin{figure*}

\begin{mathpar}
\footnotesize
\inferrule[S-Closure-Base]{
}{
    \DExp \SssStar \DExp
}

\inferrule[S-Closure-Step]{
    \DExp \Sss \DExp' \\
    \DExp' \SssStar \DExp''
}{
    \DExp \SssStar \DExp''
}

\end{mathpar}

\caption{Judgments for the transitive closure of the small step semantics.}
\label{fig:smallstepclosure}
\end{figure*}

\begin{figure}
\begin{mathpar}
\footnotesize
\inferrule[Type-Ctx-Empty-WF]{
}{
    \Wf{\cdot}
}

\inferrule[Type-Ctx-One-WF]{
    \Wf{\TypeCtx} \\ \vdash \HasType[\Elim][\DType]
}{
    \Wf{\TypeCtx, \HasType[\Elim][\DType]}
}

\end{mathpar}
\caption{Judgments for the well-formedness of a type or focusing context.}
\label{fig:typectxwf}
\end{figure}

%
%





\begin{figure}

\begin{mathpar}
\footnotesize
\boxed{\Vars} \\

\Vars[\cdot] = \{\}

\Vars[\TypeCtx, \HasType[x][\DType]] = \{\HasType[x][\DType]\} \cup \Vars[\TypeCtx]

\Vars[\TypeCtx, \HasType[\DExp][\DType]] = \Vars[\TypeCtx]~where~\DExp \neq x
\end{mathpar}

\caption{Rules for the variable extraction relation, which produces the set of all type
         bindings of variables
         from a context that might otherwise contain arbitrary elimination forms.}
\label{fig:varextraction}
\end{figure}

\subsection{Properties}

\subsubsection*{Overview}

This section describes a number of useful properties of the type-directed synthesis system
with products and focusing.  The accompanying proofs and lemmas appear in Appendix A.  Note that,
since none of the synthesis or focusing judgments modify the constructor context (\CtorCtx), it
has been elided from the following theorems for readability; for our purposes,
it is assumed to be a fixed and globally available entity.

\subsubsection*{Theorem 2.2: Progress of Focusing}

\begin{tabbing}
\noindent $\FocusCtx_1 = \cdot$ \hspace{1em} \textsc{or} \hspace{1em}
\IFocus[
    \RefineCtxM[\TypeCtx_1][\AuxCtx_1][\FocusCtx_1][\WorldCtx_1]
  ][\RefineCtxM[\TypeCtx_2][\AuxCtx_2][\FocusCtx_2][\WorldCtx_2]
  ]
\end{tabbing}

\noindent Whenever the focusing context is non-empty, a focusing judgment can be
applied.

\subsubsection*{Theorem 2.3: Preservation of Focusing}
\begin{tabbing}
\textsc{If} \hspace{1.5em} \=
               \Wf{\TypeCtx_1}, \hspace{.5em} \Wf{\AuxCtx_1},  \hspace{.5em} \Wf{\FocusCtx_1}
            \\ \textsc{and} \>
            \IFocus[
                    \RefineCtxM[\TypeCtx_1][\AuxCtx_1][\FocusCtx_1][\WorldCtx_1]
                  ][\RefineCtxM[\TypeCtx_2][\AuxCtx_2][\FocusCtx_2][\WorldCtx_2]]
            \\ \textsc{then} \>
            \Wf{\TypeCtx_2}, \hspace{.5em} \Wf{\AuxCtx_2},  \hspace{.5em} \Wf{\FocusCtx_2} 
\end{tabbing}

\noindent The application of a focusing judgment preserves the well-formedness of \TypeCtx, \AuxCtx, and
\FocusCtx.  The judgments for well-formedness are in Figure \ref{fig:typectxwf}.

\subsubsection*{Theorem 2.5: Termination of Focusing}
\begin{tabbing}
\IFocusStar[
    \RefineCtxM[\TypeCtx_1][\AuxCtx_1][\FocusCtx_1][\WorldCtx_1]
  ][\RefineCtxM[\TypeCtx_2][\AuxCtx_2][\cdot][\WorldCtx_2]
  ]
\end{tabbing}

\noindent The focusing process always terminates.  The judgments for the transitive closure of
focusing are in Figure \ref{fig:focusclosure}.

\subsubsection*{Theorem 3.1: Focusing is Deterministic}

\begin{tabbing}
\textsc{If} \hspace{2em} \=
\IFocusStar[
    \RefineCtxM[\TypeCtx_1][\AuxCtx_1][\FocusCtx_1][\WorldCtx_1]
  ][\RefineCtxM[\TypeCtx_{2a}][\AuxCtx_{2a}][\cdot][\WorldCtx_2]
  ]
    \\ \textsc{and} \>
\IFocusStar[
    \RefineCtxM[\TypeCtx_1][\AuxCtx_1][\FocusCtx_1][\WorldCtx_1]
  ][\RefineCtxM[\TypeCtx_{2b}][\AuxCtx_{2b}][\cdot][\WorldCtx_2]
  ]
    \\ \textsc{then} \>
$\TypeCtx_{2a} = \TypeCtx_{2b}$ \hspace{2em} \textsc{and} \hspace{2em}
$\AuxCtx_{2a} = \AuxCtx_{2b}$
\end{tabbing}
\noindent The focusing process proceeds deterministically.

\subsubsection*{Theorem 3.2: Order of Focusing}

Define the judgments \EGuessN~and \RefineN~
to be identical to those in Figure \ref{fig:productsrules} except that \FocusCtx~need not
be empty for any synthesis rule to be applied.

\begin{tabbing}
\RefineM \= \hspace{2em} \textsc{iff} \hspace{2em} \RefineN \\
\EGuessM \> \hspace{2em} \textsc{iff} \hspace{2em} \EGuessN 
\end{tabbing}

\noindent If we make the proof search process more non-deterministic by permitting
arbitrary interleavings of focusing and synthesis, we do not affect the set of
expressions that can be generated.  This more lenient process makes
many of the other theorems simpler to prove.

\subsubsection*{Admissibility of Focusing}

Define the judgments \EGuessOM[\TypeCtx][\DType][\Elim]
and \RefineOM~
to be identical to those in Figure \ref{fig:productsrules} without focusing. Instead of focusing
to project on tuples, we introduce the \textsc{EGuess-Proj} rule as below:
\begin{mathpar}
\inferrule*[right=EGuess-Proj]{
   \EGuessOM[\TypeCtx, \HasType[\Elim_1][\TTuple{\DType_1 \TStar ... \TStar \DType_m},~
             \HasType[\EProj[1][\Elim_1]][\DType_1],~...,~\HasType[\EProj[m][\Elim_1]][\DType_m]]
           ][\DType][\Elim]
}{
   \EGuessOM[\TypeCtx, \HasType[\Elim_1][\TTuple{\DType_1 \TStar ... \TStar \DType_m}]][\DType][\Elim]
}
\end{mathpar}

\subsubsection*{Theorem 4.1: Soundness of Focusing}

\begin{tabbing}
\noindent \textsc{If} \hspace{2em}
\RefineM[\TypeCtx][\cdot][\cdot][\cdot][\DType][\Intro] \=
\hspace{2em} \textsc{then} \hspace{2em}
\RefineOM[\TypeCtx][\cdot][\DType][\Intro] \\
\noindent \textsc{If} \hspace{2em}
\EGuessM[\TypeCtx][\cdot][\cdot][\DType][\Intro] \>
\hspace{2em} \textsc{then} \hspace{2em}
\EGuessOM[\TypeCtx][\DType][\Intro]
\end{tabbing}

\noindent Any judgment that can be proven in the system with focusing can be proven in the system
without focusing.  That is, the system with focusing is no more powerful than the system
without focusing.

\subsubsection*{Theorem 4.2: Completeness of Focusing}

\begin{tabbing}
\noindent \textsc{If} \hspace{2em}
\RefineOM[\TypeCtx][\cdot][\DType][\Intro] \=
\hspace{2em} \textsc{then} \hspace{2em}
\RefineM[\TypeCtx][\cdot][\cdot][\cdot][\DType][\Intro] \\
\noindent \textsc{If} \hspace{2em}
\EGuessOM[\TypeCtx][\DType][\Intro] \>
\hspace{2em} \textsc{then} \hspace{2em}
\EGuessM[\TypeCtx][\cdot][\cdot][\DType][\Intro]
\end{tabbing}

\noindent Any judgment that can be proven in the system without focusing can be proven in the system
with focusing.  That is, the system with focusing is at least as powerful as the system
without focusing.  Together, Theorems 4.1 and 4.2 demonstrate that the system with focusing
is exactly as powerful as the system without focusing

\subsubsection*{Note 4.3: Pushing Examples Through Elimination Forms}

Observe that we can push examples through the tuple focusing judgments but not the
\textsc{EGuess-Proj} judgment.  The sole distinction between these two judgments is
where they may be used in the synthesis process: \textsc{EGuess-Proj} must be used during
the elimination-guessing phase while focusing may take place at any point in the synthesis
process.  This discrepancy demonstrates that it is possible, at least in some cases, to make
use of example information when synthesizing elimination forms.  Since guessing elimination
forms requires raw term enumeration, any constraints on this process would yield significant
performance improvements.  I leave further investigation of this behavior to future work.

\subsubsection*{Theorem 5.1: Soundness of Synthesis Judgments}

\begin{tabbing}
\textsc{If} \hspace{2em} \= \RefineM[\TypeCtx][\AuxCtx][\FocusCtx][\Overbar{n}{\World}] \\
\textsc{then} \> \HasTypeCtx[\Vars[\TypeCtx, \AuxCtx, \FocusCtx]][\Intro][\DType]
\hspace{1em} \textsc{and} \hspace{1em} $\forall~i \in n,~\ExamCtx_i(\Intro) \SssStar \DExam_i$ \\
\textsc{If} \> \EGuessM[\TypeCtx][\AuxCtx][\FocusCtx][\DType][\Elim] \\
\textsc{then} \> \HasTypeCtx[\Vars[\TypeCtx, \AuxCtx, \FocusCtx]][\Elim][\DType]
\end{tabbing}

\noindent The type-directed synthesis system will always produce expressions with the desired
type that obey the input-output examples.  The typing judgments, small-step semantics, and 
variable extraction relation are defined in Figures \ref{fig:types}, \ref{fig:smallstep}, and \ref{fig:varextraction}.

\subsubsection*{Completeness of Synthesis Judgments}

Stating a completeness theorem for type-directed synthesis, let alone proving it,
remains an open question. 
I therefore do not attempt to prove completeness here.

%% file: products-grammar.tex

\begin{tabular}{l l l}

\DType &
\GEq \TBase \GBar \TFunc{\DType_1}{\DType_2} \GBar \hlm{\TUnit}
            \GBar \hlm{\TTuple{\DType_1 \TStar ... \TStar \DType_m}}&
(Types: base type, function type, \hl{unit type}, and \hl{product type})\\

\DVal &
\parbox[t]{5cm}{
\GEq \hlm{\EUnit} \GBar \hlm{\ETupleRange{\DVal}} \\
\phantom{1em}\GBar \ECtor[\DCtor][\hlm{\DVal}] \\
\phantom{1em}\GBar \EFix \\
\phantom{1em}\GBar \Partial
}
&
\parbox[t]{10cm}{
(Values: \hl{unit}, \hl{tuples}, \hl{single-arity} constructors, \\ functions, and partial functions)
}\\

\Partial &
\GEq \Overbar{i \in m}{\DVal_i \Rightarrow \DExam_i} &
(Partial functions: a finite map from inputs to outputs)\\

\DExam &
\GEq \hlm{\EUnit} \GBar \hlm{\ETupleRange{\DExam}} \GBar \ECtor[\DCtor][\hlm{\DExam}] \GBar \Partial &
(Examples)\\

\DExp &
\parbox[t]{5cm}{
\GEq \DVar \GBar \Partial \\
\phantom{1em}\GBar \hlm{\EUnit} \GBar \hlm{\ETuple} \GBar \hlm{\EProj} \\
\phantom{1em}\GBar \EFix \GBar \EApp \\
\phantom{1em}\GBar \ECtor[\DCtor][\hlm{\DExp}] \\
\phantom{1em}\GBar \EMatch[\DExp][\Overbar{i \in l}{\DCtor_i~\hlm{\DVar} \rightarrow \DExp_i}]
}
&
(Expressions: \hl{projection} is newly added)\\

& & \\

\Elim &
\GEq \DVar \GBar \EApp[\Elim][\Intro] \GBar \hlm{\EProj[k][\Elim]} &
(Elimination forms: \hl{projection} is newly added)\\

\Intro &
\parbox[t]{5cm}{
\GEq \Elim \\
\phantom{1em}\GBar \hlm{\EUnit} \GBar \hlm{\ETupleRange{\Intro}} \\
\phantom{1em}\GBar \ECtor[\DCtor][\hlm{\Intro}] \\
\phantom{1em}\GBar \EFix[f][x][\DType_1][\DType_2][\Intro] \\
\phantom{1em}\GBar \EMatch[\Elim][\Overbar{i \in l}{\DCtor_i~\hlm{\DVar} \rightarrow \Intro_i}]
}
 &
(Introduction forms: \hl{unit} and \hl{tuples} are newly added)\\

& & \\

\CtorCtx &
\GEq \GEmp \GBar \CtorCtx, \HasType[\DCtor][\CtorType{\hlm{\DType}}] &
(Constructor context: a map from constructor names to types) \\

\TypeCtx, \hlm{\AuxCtx, \FocusCtx} &
\GEq \GEmp \GBar \TypeCtx, \HasType[\hlm{\Elim}][\DType] &
(Type context: bindings of \hl{elimination forms} to types) \\

\ExamCtx &
\GEq \GEmp \GBar \ExamCtx, \Refines[\hlm{\Elim}][\DExam] &
(Example context: \hl{elimination forms} refined by examples) \\

\WorldOne &
\GEq \World &
(A single world: an example context and result example) \\

\WorldCtx &
\GEq \GEmp \GBar \WorldCtx, \WorldOne &
(A set of worlds) \\

\end{tabular}

%% file: focusing-rules.tex

\begin{gather*}
\boxed{\IFocus[
    \RefineCtx[\CtorCtx_1][\TypeCtx_1][\AuxCtx_1][\FocusCtx_1][\WorldCtx_1]
  ][\RefineCtx[\CtorCtx_2][\TypeCtx_2][\AuxCtx_2][\FocusCtx_2][\WorldCtx_2]
  ]
} \\ \\
\inferrule[Focus-Unit]{
}{
    \IFocus[
        \RefineCtx[\CtorCtx][\TypeCtx][\AuxCtx][\FocusCtx, \HasType[\Elim][\TUnit]][\WorldCtx]
      ][\RefineCtx[\CtorCtx][\TypeCtx, \HasType[\Elim][\TUnit]][\AuxCtx][\FocusCtx][\WorldCtx]]
} ~~~~~~~~~~
\inferrule[Focus-Base]{
}{
    \IFocus[
        \RefineCtx[\CtorCtx][\TypeCtx][\AuxCtx][\FocusCtx, \HasType[\Elim][\TBase]][\WorldCtx]
      ][\RefineCtx[\CtorCtx][\TypeCtx, \HasType[\Elim][\TBase]][\AuxCtx][\FocusCtx][\WorldCtx]]
} \\ \\
\inferrule[Focus-Fun]{
}{
    \IFocus[
        \RefineCtx[\CtorCtx][\TypeCtx][\AuxCtx
                 ][\FocusCtx, \HasType[\Elim][\TFunc{\DType_1}{\DType_2}]][\WorldCtx]
      ][\RefineCtx[\CtorCtx][\TypeCtx, \HasType[\Elim][\TFunc{\DType_1}{\DType_2}]
                 ][\AuxCtx][\FocusCtx][\WorldCtx]]
} \\ \\
\inferrule[Focus-Tuple]{
}{
      \RefineCtx[\CtorCtx
            ][\TypeCtx
            ][\AuxCtx
            ][\FocusCtx, \HasType[\Elim][\TTuple{\DType_1 \TStar ... \TStar \DType_m}]
            ][\Overbar{i \in n}{\World[\ExamCtx_i,
                                        \Refines[\Elim][\ETuple[\Intro_{(i, 1)}, ..., \Intro_{(i, m)}]]
                                      ][\Exam_i]
                                }
            ] 
    \Longrightarrow \\
    \RefineCtx[\CtorCtx
            ][\TypeCtx
            ][\AuxCtx, \HasType[\Elim][\TTuple{\DType_1 \TStar ... \TStar \DType_m}]
            ][\FocusCtx, \Overbar{j \in m}{\HasType[\EProj[j][\Elim]][\DType_j]}
            ][\Overbar{i \in n}{\World[\ExamCtx_i,
                                       \Overbar{j \in m}{\Refines[\EProj[j][\Elim]][\Intro_{(i, j)}]}
                                     ][\Exam_i]
                               }
            ]
}
\end{gather*}

%% file: products-rules.tex
\begin{gather*}
\boxed{\Refine[\CtorCtx][\TypeCtx][\hlm{\AuxCtx}][\hlm{\FocusCtx}]} \\ \\
\inferrule[IRefine-Guess]{
    \EGuess[\CtorCtx][\TypeCtx][\AuxCtx][\cdot][\TBase][\Elim] \\
    \forall i \in n,~ \ExamCtx_i(\Elim) \SssStar \Exam_i
}{
    \Refine[\CtorCtx][\TypeCtx][\AuxCtx][\cdot][\Overbar{i \in n}{\World[\ExamCtx_i][\Exam_i]}][\TBase][\Elim]
} ~~~~~~~~~~~~~~~~~~~~~~~~~
\hlm{
\inferrule[IRefine-Unit]{
}{
    \Refine[\CtorCtx][\TypeCtx][\AuxCtx][\cdot][\Overbar{i \in n}{\World[\ExamCtx_i][\EUnit]}][\TUnit][\EUnit]
}
} \\ \\
\inferrule[IRefine-Fix]{
    \Refine[\CtorCtx
          ][\TypeCtx
          ][\AuxCtx][\HasType[\DFun][\TFunc{\DType_1}{\DType_2}], \HasType[\DVar][\DType_1]
          ][\Overbar{(i, k) \in (n, m_i)}{\World[\ExamCtx_i,
                                             \Refines[f][\Overbar{j \in m_i}{\DVal_{(i, j)}
                                                         \Rightarrow \DExam_{(i, j)}}],
                                             \Refines[x][\DVal_{(i, k)}]
                                           ][\DExam_{(i, k)}]}
          ][\DType_2
          ][\Intro]
}{
    \Refine[\CtorCtx
          ][\TypeCtx
          ][\AuxCtx][\cdot
          ][\Overbar{i \in n}{\World[\ExamCtx_i
                                   ][\Overbar{j \in m_i}{\DVal_{(i, j)} \Rightarrow \DExam_{(i, j)}}]}
          ][\TFunc{\DType_1}{\DType_2}
          ][\EFix[f][x][\DType_1][\DType_2][\Intro]]
} \\ \\
\hlm{
\inferrule[IRefine-Ctor]{
    \HasType[\DCtor][\TFunc{\DType}{\TBase}] \in \CtorCtx \\
    \Refine[\CtorCtx][\TypeCtx][\AuxCtx][\cdot
          ][\Overbar{i \in n}{\World[\ExamCtx_i][\Exam_i]}][\DType][\Intro]
}{
    \Refine[\CtorCtx][\TypeCtx][\AuxCtx][\cdot
          ][\Overbar{i \in n}{\World[\ExamCtx_i][\ECtor[\DCtor][\Exam_i]]}][\TBase][\ECtor[\DCtor][\Intro]]
}
} ~~~~~~
\hlm{
\inferrule[IRefine-Tuple]{
    \forall i \in m,~ \Refine[\CtorCtx][\TypeCtx][\AuxCtx][\cdot
                            ][\Overbar{j \in n}{\World[\ExamCtx_j][\Exam_{(i, j)}]}
                            ][\DType_i][I_i]
}{
    \Refine[\CtorCtx][\TypeCtx][\AuxCtx][\cdot
          ][\Overbar{j \in n}{\World[\ExamCtx_j
                                   ][\ETuple[\Exam_{(1, j)}, ..., \Exam_{(m, j)}]]
                             }
          ][\TTuple{\DType_1 \TStar ... \TStar \DType_m}
          ][\ETupleRange{\Intro}]
}
} \\ \\
\hlm{
\inferrule[IRefine-Match]{
    \EGuess[\CtorCtx][\TypeCtx][\AuxCtx][\cdot][\TBase][\Elim] \\
    \forall j \in m, ~\HasType[\DCtor_j][\TFunc{\DType_j}{\TBase}] \in \CtorCtx \\
    \forall j \in m, ~\Refine[\CtorCtx][\TypeCtx][\AuxCtx][\HasType[\DVar][\DType_j]
                             ][\{\World[\ExamCtx_i, \Refines[\DVar][\DVal]][\Exam_i]~\big|~
                               \ExamCtx_i(\Elim) \SssStar \DCtor_j~\DVal\}
                              ][\DType][\Intro_j]
}{
    \Refine[\CtorCtx][\TypeCtx][\AuxCtx][\cdot
          ][\Overbar{i \in n}{\World[\ExamCtx_i][\Exam_i]}][\DType
          ][\EMatch[\Elim][\Overbar{j \in m}{\DCtor_j~\DVar \rightarrow \Intro_j}]]
}
} \\ \\
\hlm{
\inferrule[IRefine-Focus]{
    \IFocus[
        \RefineCtx[\CtorCtx_1][\TypeCtx_1][\AuxCtx_1][\FocusCtx_1][\WorldCtx_1]
    ][  \RefineCtx[\CtorCtx_2][\TypeCtx_2][\AuxCtx_2][\FocusCtx_2][\WorldCtx_2]
    ]
    \\
    \Refine[\CtorCtx_2][\TypeCtx_2][\AuxCtx_2][\FocusCtx_2][\WorldCtx_2][\DType][\Intro]
}{
    \Refine[\CtorCtx_1][\TypeCtx_1][\AuxCtx_1][\FocusCtx_1][\WorldCtx_1][\DType][\Intro]
}
}
\end{gather*}


\begin{gather*}
\boxed{\EGuess[\CtorCtx][\TypeCtx][\hlm{\AuxCtx}][\hlm{\FocusCtx}][\DType][\Elim]} \\ \\
\hlm{
\inferrule[EGuess-Ctx]{
    \HasType[\Elim][\TBase] \in \TypeCtx
}{
    \EGuess[\CtorCtx][\TypeCtx][\AuxCtx][\cdot][\TBase][\Elim]
}
} ~~~~~~~~~~~~~~~~~~~~~
\hlm{
\inferrule[EGuess-App]{
    \Refine[\CtorCtx][\TypeCtx, \HasType[\Elim_1][\TFunc{\DType_1}{\DType_2}]
          ][\AuxCtx][\cdot][\cdot][\DType_1][\Intro_1] \\
    \EGuess[\CtorCtx][\TypeCtx, \HasType[\Elim_1][\TFunc{\DType_1}{\DType_2}]
          ][\AuxCtx][\HasType[\EApp[\Elim_1][\Intro_1]][\DType_2]][\DType][\Elim]
}{
    \EGuess[\CtorCtx][\TypeCtx, \HasType[\Elim_1][\TFunc{\DType_1}{\DType_2}]
          ][\AuxCtx][\cdot][\DType][\Elim]
}
} \\ \\
\hlm{
\inferrule[EGuess-Focus]{
    \IFocus[
        \RefineCtx[\CtorCtx_1][\TypeCtx_1][\AuxCtx_1][\FocusCtx_1][\cdot]
    ][  \RefineCtx[\CtorCtx_2][\TypeCtx_2][\AuxCtx_2][\FocusCtx_2][\cdot]
    ]
    \\
    \EGuess[\CtorCtx_2][\TypeCtx_2][\AuxCtx_2][\FocusCtx_2][\DType][\Elim]
}{
    \EGuess[\CtorCtx_1][\TypeCtx_1][\AuxCtx_1][\FocusCtx_1][\DType][\Elim]
}
}
\end{gather*}

%% file: evaluation.tex
\section{Evaluation}

\subsection{Implementation Changes}

The prototype implementation of the type-directed synthesis system consists of
approximately 3,000 lines of OCaml.  The structure of the implementation largely
mirrors that of the synthesis judgments, with type and constructor contexts and
recursive proof search for a program that satisfies the input-output examples.
The implementation deviates only when searching for elimination forms, during which
it constructs terms bottom-up instead of the less-efficient top-down guessing suggested by
the synthesis judgments.

To add tuples, I altered between 1,500 and 2,000 lines of code.  Notable changes included:
\begin{itemize}
\item Modifying the parser and type-checker to add tuples, projection, and single-argument
      constructors.
\item Entirely rewriting the type context to store elimination forms and perform focusing.
\item Updating the synthesis process to conform to the new judgments.
\item Adding a unit type and value throughout the system.
\item Extending the pattern-matching language to include mechanisms for recursively
      matching on the tuples embedded within constructors. 
\item Refactoring, simplification, modularization, dead code elimination, and other cleanup
      to prepare the system for future extensions.
\end{itemize}

\subsection{Performance Evaluation}

\begin{figure}
\begin{subfigure}[b]{.5\textwidth}
\footnotesize
\begin{tabular}{|l|r|r|}
\hline
\textbf{Program}         & \textbf{Examples} & \textbf{Time (s)} \\ \hline
\texttt{make\_pair}      & 2                 & 0.001             \\ \hline 
\texttt{make\_triple}    & 2                 & 0.003             \\ \hline 
\texttt{make\_quadruple} & 2                 & 0.003             \\ \hline 
\texttt{fst}             & 2                 & 0.003             \\ \hline 
\texttt{snd}             & 2                 & 0.003             \\ \hline 
\texttt{unzip (bool)}    & 7                 & 0.005             \\ \hline
\texttt{unzip (nat)}     & 7                 & 0.006             \\ \hline
\texttt{zip (bool)}      & 28                & 1.183             \\ \hline
\texttt{zip (nat)}       & 28                & 2.879             \\ \hline

\end{tabular}

\caption{The number of examples and seconds necessary to synthesize various canonical programs that
         make use of tuples and projection.}
\label{fig:raweval}
\end{subfigure}
~~~~~
\begin{subfigure}[b]{.5\textwidth}
\footnotesize
\begin{tabular}{|l|r|r|}
\hline
\textbf{Program}                & \textbf{Before (s)} & \textbf{After (s)} \\ \hline
\texttt{bool\_band}             & 0.006               & 0.007              \\ \hline
\texttt{list\_append}           & 0.016               & 0.024              \\ \hline
\texttt{list\_fold  }           & 0.304               & 0.183              \\ \hline
\texttt{list\_filter}           & 2.943               & 0.657              \\ \hline
\texttt{list\_length}           & 0.003               & 0.003              \\ \hline
\texttt{list\_map}              & 0.001               & 0.039              \\ \hline

\texttt{list\_nth}              & 0.061               & 0.072              \\ \hline
\texttt{list\_sorted\_insert}   & 91.427              & 10.600             \\ \hline
\texttt{list\_take}             & 0.197               & 0.083              \\ \hline
\texttt{nat\_add}               & 0.016               & 0.011              \\ \hline
\texttt{nat\_max}               & 0.037               & 0.065              \\ \hline
\texttt{nat\_iseven}            & 0.005               & 0.006              \\ \hline
\texttt{tree\_binsert}          & 0.762               & 0.865              \\ \hline
\texttt{tree\_nodes\_at\_level} & 1.932               & 0.812              \\ \hline
\texttt{tree\_preorder}         & 0.081               & 0.076              \\ \hline
\end{tabular}

\caption{A comparison of the synthesis performance between the system as described in the original
         type-directed synthesis paper (Before) and the updated version of the system that
         includes the changes described in this thesis (After).}
\label{fig:compeval}
\end{subfigure}
\caption{Performance evaluation of the prototype implementation.}
\end{figure}

\paragraph{Raw Performance.}

Figure \ref{fig:raweval} summarizes the performance of the prototype implementation on a number of
canonical programs that make use of tuples and projection.  Programs that did not require recursive
calls, like \texttt{make\_pair} and \texttt{fst}, were synthesized instantly.
The list \texttt{unzip} function was likewise generated in a small fraction of a second,
but required a relatively large number of examples.  The synthesizer would
overfit if any example was discarded.  The list \texttt{zip} function took the longest to
synthesize and required far more examples.  The reason for this behavior is that \texttt{zip}
must handle many more cases than \texttt{unzip}, such as the possibility that the two lists
are of different lengths.

The \texttt{unzip} and \texttt{zip} programs were tested twice: once each with lists comprising boolean
and natural number values.  The natural number case predictably took longer, since the
synthesizer likely spent a portion of its time generating programs that
manipulate the values stored in the lists as opposed to the lists themselves.
Since there are a finite number of boolean values,
the search space of these faulty programs was much smaller.  With polymorphic lists, this
behavior could be eliminated completely.

\paragraph{Performance Regression.}

The changes made in the process of adding tuples affected the entire synthesis system.  Even programs
that do not directly make use of tuples could take longer to synthesize.  All multi-argument
constructors, including \texttt{Cons}, now require tuples when created and projection when destroyed.
No-argument constructors, which formerly served as the base cases of the synthesis
process, now store values of unit type.
Whenever a variable is added to the context, it goes through the process of focusing before
it can be used.  While small, these modifications add overhead that impacts any program
we might attempt to synthesize.

Figure \ref{fig:compeval} contains a performance comparison of the system as described in the original
paper with the updated version that includes tuples.  The running times were generated
using the same computer on the same date; they are not lifted directly from the
type-directed synthesis paper~\cite{tds}.  The selected programs are a subset
of those tested in the original paper. Overall, it appears that there has been little tangible
change in the performance of the system.  Most examples saw at most slight fluctuations.  Several
of the longer-running examples, however, experienced dramatic performance improvements, which might
have resulted from changes made to the memoization behavior of the system when
rewriting the type context.

%% file: related.tex
\section{Related Work}

\subsection{Solver-free Synthesis}
\label{sec:solver-free}

\paragraph{StreamBit.}

Among the first modern synthesis techniques was sketching~\cite{sketch1}, developed by
Armando Solar-Lezama, which allows users
to specify an incomplete program (a \emph{sketch}) that is later filled in by a
synthesizer.  Although this line of work evolved into a general-purpose, solver-aided synthesis
language (see Section \ref{sec:solver-aided}), the initial prototype was designed for
programs that manipulate streams of bits (\emph{bitstreaming} programs), a notable
example of which is a cryptographic cipher.  In the system's language, StreamBit,
users specify composable transformers, splits, and joins on streams of bits that together
form a program.  The paper states that StreamBit helped programmers design optimized ciphers
that were several times faster than existing versions written in C.

The process of synthesis in StreamBit involves two steps.  First, a user writes a correctness
specification in the form of a program that implements the desired functionality.  A user then
writes a sketch that describes some, but not all, of the program to be synthesized.  The remaining
integer and boolean holes are filled by the synthesizer (using mathematical
transformations and linear algebra) to create a program whose behavior is equivalent to the
specification.  The theory behind this design is that it is easy to write the naive, unoptimized
versions of bitstreaming programs that serve as specifications but difficult to manage the low-level
intricacies of optimized variants.  Sketching allows a human to describe the algorithmic insights
behind possible optimizations while leaving the synthesizer to perform the complex
task of working out the details.

Sketching also permits a programmer to determine that a particular syntactic template is
\emph{not} a viable solution should the synthesizer fail to generate a program.  (The idea of
discovering an algorithm in top-down fashion by testing syntactic templates for viability with
the aid of a synthesizer or - more broadly - a solver is discussed at length in \cite{angelic}.)

\paragraph{Flash Fill.}

Designed by Sumit Gulwani, Flash Fill~\cite{flashfill} synthesizes string
transformations using input-output examples.
The system is featured in Excel 2013, where it generalizes user-completed data transformations
on a small set of examples into a formula that works for the larger spreadsheet.  The synthesizer
behind Flash Fill relies on a domain-specific language with constructs for substring manipulation,
regular expression matching, loops, and conditionals.

Flash Fill works by examining each example in turn.  It uses a graph to store the set of all possible
transformations that could convert each input into its corresponding output. By taking
the intersection of these graphs among all examples, Flash Fill synthesizes a transformation
that works in the general case.  This process relies on a variety of string manipulation
algorithms and relations between transformations rather than a solver.

Flash Fill epitomizes several key synthesis design concepts.  Like type-directed
synthesis, Flash Fill derives a program from input-output examples, which are more
convenient for programmers and non-programmers alike, rather than a complete
specification.  It uses a compact representation - namely a transformation graph - to
capture a large number of possible programs in a small amount of storage.  Finally, it
synthesizes programs over a specialized, domain-specific language whose search space
is smaller than that of a general-purpose language, reducing the difficulty of the overall
synthesis task.

\paragraph{Escher.}

The closest relative to type-directed synthesis is Gulwani's general-purpose synthesis
system, Escher~\cite{escher}.  Like type-directed synthesis, Escher aims to generate
general-purpose, recursive programs from input-output examples, although Escher assumes
the existence of an oracle that can be queried when the synthesizer needs additional
guidance.  Gulwani notes that
Escher's synthesis process is ``generic'' in that it works over an arbitrary domain
of instructions, although he only demonstrates the system on a language with
algebraic datatypes.

In contrast to type-directed synthesis, Escher uses bottom-up \emph{forward search},
generating programs of increasing size out of instructions, recursive calls
to the function being synthesized, input variables, and other language atoms.
Escher maintains a \emph{goal graph}, which initially comprises a single node
that stores the set of outputs a synthesized program would need to produce to satisfy the
examples and solve the synthesis problem.  When a generated program $p$ satisfies some
(but not all) of the outputs in a goal node, Escher creates a conditional that
partitions the program's control flow, directing the satisfied cases to $p$.
Doing so creates two new goal nodes: one for the conditional's guard and the other to handle
the cases that $p$ failed to satisfy.  When no more unsatisfied goal nodes remain,
Escher has synthesized a program.

\subsection{Solver-aided Synthesis}
\label{sec:solver-aided}

\paragraph{Syntax-guided Synthesis.}

Published in 2013 and co-authored by many leading researchers in solver-aided synthesis
(including Rastislav Bodik, Sanjit Seshia,
Rishabh Singh, Armando Solar-Lezama, Emina Torlak, and others),
\emph{Syntax-Guided Synthesis} (SyGuS)~\cite{sygus} represents an attempt to codify a
``unified theory'' of synthesis.  The authors seek to create a standardized
input format for synthesis queries, similar to those developed for SMT and SAT solvers.
Doing so for SMT and SAT solvers has led to shared suites of benchmarks,
objective performance comparisons between solvers, and annual solver competitions, an effort
credited with encouraging researchers to advance the state of the art.  More
importantly, a common interface means that higher-level applications can
call off-the-shelf solvers, abstracting away solvers as subroutines.  When solver
performance improves or better solvers are developed, applications instantly benefit.
The SyGuS project seeks to create a similar environment for synthesizers~\cite{sygus-website}.

The common synthesis format encodes three pieces of information: a \emph{background theory} in
which the synthesis problem will be expressed, a \emph{correctness specification} dictating the
requirements of an acceptable solution, and the \emph{grammar} from which candidate
programs may be drawn.

The paper includes a thorough taxonomy of modern synthesis techniques.  It distinguishes between
\emph{deductive synthesis}, in which a synthesizer creates a program by proving its logical
specification, and \emph{inductive synthesis}, which derives a program from input-output examples.
It also describes a popular inductive synthesis algorithm, \emph{counterexample-guided inductive
synthesis} (CEGIS).  In CEGIS, a solver repeatedly generates programs that it attempts to
verify against a specification.  Whenever verification fails, the synthesis procedure can use
the verifier's counterexample as an additional constraint in the program-generation procedure.  If no
new program can be constructed, synthesis fails; if no counterexample can be found, synthesis
succeeds.  In practice, most synthesis problems require only a few CEGIS steps to generate
the counterexamples necessary to discover a solution.

Finally, the authors present an initial set of benchmarks in the common synthesis format and
a performance comparison of existing synthesis systems.

In the sections that follow, I detail several lines of research that fit into the framework
of solver-aided syntax-guided synthesis.

\paragraph{Rosette.}

Rosette, first presented in \cite{rosette1} by Emina Torlak, is a framework for
automatically enabling solver-aided queries in domain-specific languages (DSLs).  Rosette
compiles a subset of Racket into constraints (although the full language can be
reduced to this subset), thereby enabling the same for any DSL embedded in Racket.
These constraints are then supplied to a solver, which can facilitate fault-localization,
verification, angelic non-determinism, and synthesis.  To allow DSLs to interact with
the solver directly, Rosette includes first-class \emph{symbolic} constants, which
represent holes whose values are constrained by assertions and determined by calls
to the solver.

Rosette's synthesis procedure fills a hole in a partially-complete program given
a bounded-depth grammar (a ``sketch'') from which solutions are drawn.
Correctness specifications come in the form of assertions or another program
that implements the desired functionality.  Rosette compiles the grammar into a nondeterministic program
capturing every possible AST that the grammar might produce.  Nondeterministic choices are
compiled as symbolic values.  The solver determines the nondeterministic choices necessary
to fill the hole in a way that conforms to the specification, thereby selecting an
appropriate AST and solving the synthesis problem.

In \cite{rosette2}, Torlak details the symbolic compiler underlying Rosette.
The paper includes a textbook-quality discussion of two possible techniques for compiling
Racket into constraints: \emph{symbolic execution} and \emph{bounded model checking}.  In
symbolic execution, the compiler must individually explore every possible code path of the program,
branching on loops and 
conditionals.  This approach concretely executes as much of the program as possible but produces
an exponential-sized output.  In contrast, bounded model checking joins program states back
together after branching, creating new symbolic $\phi$-values representing values merged from
different states.  The
resulting formula is smaller but contains fewer concrete values.  Both approaches require
bounded-depth loop unrolling and function inlining to ensure termination of the compilation
procedure.

Torlak adopts a hybrid strategy that draws on the strengths of both techniques:
Rosette uses bounded model checking but, when merging, first joins values like lists
structurally rather than symbolically, preserving opportunities
for concrete evaluation.  The compiler assumes that the caller has
implemented some form of termination guarantee, such as structural recursion on an inductive
datastructure or an explicit limit on recursion depth.

\paragraph{Kaplan, Leon, and related synthesis work.}

Viktor Kuncak's Kaplan~\cite{kuncak-control} explores the design space of languages
with first-class constraints, symbolic values, and the ability to make calls to
a solver at run-time in a manner similar to Rosette, adding these features to a pure
subset of Scala.  Kaplan's constraints are functions from inputs parameters to a
boolean expression.  Calling \emph{solve} on a constraint generates concrete values
for the input parameters that cause the boolean expression to be true.  Kaplan also
includes mechanisms for iterating over all solutions to a constraint, choosing a particular solution
given optimality criteria, or lazily extracting
symbolic rather than concrete results.  Since constraints are simply functions, they can
be composed or combined arbitrarily.

Kaplan introduces a new control-flow construct: an $\mathtt{assuming...otherwise}$ block.
This structure functions like an $\mathtt{if}$-statement where the guard is a constraint rather
than a boolean expression.  If the constraint has a solution, the code in the $\mathsf{assuming}$
branch is executed; if it is unsatisfiable, the $\mathtt{otherwise}$ block executes instead.

Kuncak has also explored the opposite approach, performing synthesis entirely at compile-time
rather than run-time.   Invoking a solver at run-time, he argues,
is unpredictable and unwieldy.  Ideally, synthesis should function as a compiler service that
provides immediate, static feedback to developers and always succeeds when a synthesis problem
is feasible. In \cite{kuncak-complete}, Kuncak attempts to turn decision procedures
for various theories (i.e., linear arithmetic and sets) into corresponding,
\emph{complete} synthesis procedures.  As in Kaplan, users specify constraints, but the
synthesis engine derives a program that statically produces values for the constraint's inputs
instead of invoking a solver at run-time.

Combining the lessons of these two systems, Kuncak developed a synthesis engine for creating
general-purpose, recursive programs~\cite{kuncak-recursive}.  The synthesizer uses a hybrid
algorithm with deductive and inductive steps.  The synthesis problem is expressed as a logical
specification to which a program must conform rather than individual input-output examples.

In the deductive step, the synthesizer applies various rules for converting a specification into
a program in a fashion similar to type-directed synthesis.  Special rules are generated to
facilitate recursively traversing each available inductive datatype.

At the leaves of this deductive process, the synthesizer uses an inductive CEGIS step to choose
candidate expressions.  Like Rosette, it encodes bounded-depth ASTs that generate terms of a
particular type as nondeterministic programs.  By querying a solver for the
nondeterministic choices that satisfy the specification, the synthesizer derives a
candidate expression.  The synthesizer then attempts to verify this expression.  If verification
fails, it receives a new counterexample that it integrates into the expression-selection process.
The synthesizer increases the AST depth until some maximum value, at which point it determines
that the synthesis problem is infeasible.  As the program size grows, verification becomes expensive,
so the synthesizer uses concrete execution to eliminate candidate programs more quickly.

One notable innovation behind this system is that of \emph{abductive reasoning}, where
conditionals that partition the problem space are synthesized, creating more restrictive
path conditions that make it easier to find candidate solutions.  This method is similar
to that in Escher~\cite{escher}, which creates a conditional whenever it synthesizes a
program that partially satisfies the input-output examples.

\paragraph{Sketch.}

The synthesis technique of sketching (introduced in Section \ref{sec:solver-free} with
the StreamBit language) eventually culminated in a C-like, general-purpose programming
language, called Sketch, which synthesizes values for integer and boolean holes.  Although
Sketch's holes are limited to these particular types, users can creatively place holes
in conditional statements to force the synthesizer to choose between alternative code blocks
in a manner similar to Rosette.  Like StreamBit, a Sketch specification is a program that
implements the desired functionality.

Sketch was first introduced by Solar-Lezama in \cite{sketch2} and continues to see active
development~\cite{sketch-website}.  In \cite{sketch2}, Solar-Lezama describes both the
core Sketch language and the underlying, solver-aided synthesis procedure, CEGIS. 
Given a specification program $s$ and a partial program $p$ with input values $\vec{x}$ and
holes $\vec{h}$, Sketch's synthesis problem may be framed as the question of
whether $\exists~\vec{h},~\forall~\vec{x},~p(\vec{h}, \vec{x}) = s(\vec{x})$. Solving this
quantified boolean formula is $\Sigma_2$-complete
(i.e., at least as hard as NP-complete), making it too difficult to handle directly.

Instead, Solar-Lezama presents a two-step process for synthesizing a solution.  First, a
solver guesses values $\vec{v}$ for $\vec{h}$.  A verifier then checks whether
$\forall~\vec{x},~p(\vec{v}, \vec{x}) = s(\vec{x})$.  If the verification step succeeds, then
$\vec{v}$ is a solution to the synthesis problem; otherwise, the verifier returns a
counterexample input $\vec{c_1}$.  The solver then guesses new values $\vec{v}$
for $\vec{h}$ by integrating $\vec{c_1}$ into the formula: $\exists~\vec{h},~
p(\vec{h}, \vec{c_1}) = s(\vec{c_1})$.  If no values for $\vec{h}$ can be found, synthesis
fails; otherwise, we return to the verification step and the CEGIS loop continues.  In
general, it takes no more than a few examples to pinpoint values for $\vec{h}$.

The Sketch language presented in the original paper includes several macro-like operators
for managing holes at a higher level.  There are \emph{loop} statements that repeat a hole for a particular
number of iterations and \emph{generators} that abstract holes into recursive hole-generating functions.
Both structures are evaluated and inlined at compile-time.  Later versions of Sketch introduce
a regular expression-like syntax for describing AST grammars and assertions in place
of a specification program~\cite{sketch}.  Additional papers
add features like concurrency~\cite{sketch3}.  Solar-Lezama's dissertation collects these ideas
in a single document~\cite{sketch}.

\paragraph{Autograder.}

In an exemplary instance of using synthesis as a service, Rishabh Singh built on Sketch to
create an autograder for introductory computer science assignments~\cite{autograder}.  The
autograder relies on two premises: (1) a solution program is available and (2) common
errors are known.  A user specifies \emph{rewrite rules} consisting of possible corrections
to a student program.

The autograder compiles student programs into Sketch.  At each
location where a rewrite rule might apply, the autograder inserts a conditional statement
guarded with a hole that chooses whether to leave the previous statement in place or
take the result of a rewrite rule.  When Sketch is run on the student program with the
solution as its specification, the values for the holes determine the corrections necessary
to fix the student program.  Singh had to modify Sketch's CEGIS algorithm to also consider an
optimization constraint: the holes should be filled such that the smallest number of changes
are made to the original program.

\subsection{Theorem Proving}

Frank Pfenning's lecture notes for his course \emph{Automated Theorem Proving}~\cite{atp} are a
textbook-quality resource on techniques for efficient proof search in constructive
logic.  Building off of an introduction to natural deduction and its more wieldy analog, the
sequent calculus, Pfenning describes methods for performing forward and backward search over
first order logic.  Among strategies
presented are \emph{focusing}, in which one should eagerly apply proof steps whose effects are
invertible, and the \emph{subformula property}, which states that, in forward search, the only
necessary proof steps are those that produce a subformula of the goal statement.
The concepts portrayed in Pfenning's text, together with Gentzen's classic treatment of the
sequent calculus~\cite{gentzen}, represent the theoretical underpinnings of many of the
ideas contained in this thesis.

%% file: future-work.tex
\section{Future Work}

\subsection{Additional Language Features}

\paragraph{Records and Polymorphism.}

Although the original type-directed synthesis system included only
the simply-typed lambda calculus with small additions, this representation
already captured a useful portion of a pure subset of ML.  This thesis adds a
number of missing components, including tuples and the unit type.  The most prominent
absent features are records and polymorphism.

Records should generally resemble tuples, however their full ML specification
would require adding subtyping to the synthesis system.  The primary dividends
of integrating records would be to explore synthesis in the presence of subtyping
and to help transform the type-directed synthesis implementation from
a research prototype into a fully-featured programming tool.

Polymorphism presents a wider array of challenges and benefits.  System F-like parametric polymorphism
would require significant theoretical, syntactic, and implementation changes to the
synthesis system.  Adding the feature would, however, bring the synthesis language
far closer to ML and dramatically improve performance in some cases.

Consider,
for example, the process of synthesizing \emph{map}.  If the synthesizer knows that
\emph{map} will always be applied to a list of natural numbers, it is free use
the \texttt{S} constructor or to match on an element of the list.  If, instead, \emph{map}
were to work on a polymorphic list about whose type we know nothing, then the synthesizer
will have far fewer choices and will find the desired program faster.  This
behavior is a virtuous manifestation of Reynolds' abstraction theorem~\cite{reynolds}.

\paragraph{Solver-Aided Types.}

The existing synthesis system works only on algebraic datatypes.  A logical next step would
be to integrate true, 32-bit integers and 64-bit floating point numbers.  Doing so within the
existing framework, however, would be impossibly inefficient. 
We could represent integers as an algebraic datatype
with $2^{32}$ base cases, but the synthesizer would time out while guessing every possible number.

Instead, we might integrate with a solver equipped with the theory of bitvectors.  These tools
have proven to be quite efficient in other, first-order synthesis systems~\cite{sketch}.  By
combining the higher-order power of type-directed synthesis with the performance of modern
SMT-solvers, we could leverage the best of both approaches while bringing the system's language
closer to full ML.

\paragraph{Monads and Linear Types.}

Type-directed synthesis is far more extensible than most other synthesis frameworks:
in order to add a new language feature, we need only consult the corresponding system
of logic. A number of desirable features map to well-understood logical frameworks.
For example, monads
are captured by lax logic and resources by linear logic.  By integrating these systems of
logic into our synthesis judgments, we could generate effectful programs that raise exceptions or
read from and write to files.  These operations are at the core of many ``real'' programs but remain beyond
the capabilities of any existing synthesis system.

\subsection{Refinements and Intersections}

In a general sense, types represent (possibly-infinite) sets of expressions.  Although we
are used to the everyday notions of integers, floating-point numbers, and boolean values,
we could also craft types that contain a single item or even no elements at all.  This is
the insight behind \emph{refinement types}~\cite{refinement}, which create new, more
restrictive subtypes out of existing type definitions.  For example, we might create the type
\texttt{Nil} as a subtype of \texttt{list} that contains only the empty list and
\texttt{Cons(nat, list)} that captures all non-empty lists.  Refinements may be understood
as simple predicates on types: stating that $x$ has refinement \texttt{2} is equivalent to
the type $\{x \in \texttt{nat}~|~x = 2\}$.

The entire example grammar $\DExam$ from the type-directed synthesis system comprises a language of
refinements on the type system $\DType$.  In order to fully express partial functions as refinements,
we need one
additional operator: intersection.  The refinement \TAnds[\DExam_1, \DExam_2] requires that an
expression satisfy both $\DExam_1$ and $\DExam_2$.  For example, the \texttt{inc} function
should obey the both \TFunc{0}{1} and \TFunc{1}{2}.  Partial functions are simply
intersections of arrow refinements.

There are two benefits to characterizing examples as types.  The first is that, by lifting
examples into the type system, we are able to fully exploit the Curry-Howard isomorphism
and decades of research into automated theorem proving.  Rather than forging ahead into
unknown territory with examples, we are treading familiar ground with an especially
nuanced type system.  The second is that we can freely draw on more complex operators
from refinement type research to enrich our example language.  For instance, we might add base types
as refinements so users can express the set of all non-empty lists as discussed before, or
integrate union types to complement intersections.

Still richer refinements are possible.  One intriguing avenue is that of negation, a
refinement that functions as the set complement operator over types.
Negation allows us to capture the concept of negative examples, for example \TNot{\TFunc{0}{1}},
or to express to the synthesizer that a particular property of a program does not hold.
Adding negation to the type-directed synthesis system enables new algorithms for generating
programs, including a type-directed variant of CEGIS.  Negation represents one of many possible
specification language enhancements that refinements make possible.  We might consider adding
quantifiers over specifications or even dependent refinements.

%% file: conclusions.tex
\section{Conclusions}

Type-directed synthesis offers a new approach to an old, popular, and
challenging problem.
Where most existing systems rely on solvers to generate programs
from specifications, type-directed synthesis applies techniques from
automated theorem proving.  At the time of writing, however, type-directed
synthesis extends only to a minimal language: the simply-typed lambda
calculus with recursive functions and algebraic datatypes.

In this thesis, I have described extending type-directed synthesis
to include product types, which required rethinking both the underlying theory
and the prototype implementation.  Products simplify many of the existing rules
but invite a host of other theoretical challenges.  I leveraged the fact that
they are invertible to speed up the proof search process with focusing, which
in turn required new judgment forms and additional proofs about the behavior
of the updated rules.

The prototype implementation now efficiently synthesizes tuples and projections,
validating the new theory.  More importantly, it concretely demonstrates a key
advantage of type-directed synthesis: extending the algorithm with new syntax
is as simple as integrating the corresponding systems of logic.
Although products represent a seemingly small addition, they pave the way
for the synthesis of other, more powerful language features.

%% file: acknowledgements.tex
\section{Acknowledgements}

I would like to thank:
~\\

\noindent\textbf{My adviser, Professor David Walker}, who
has spent the past two years teaching me
nearly everything I know about programming languages and research.
I am grateful beyond measure for his patient guidance, generosity
with his time, and willingness to share his expertise. \\

\noindent\textbf{Professor Steve Zdancewic and Peter-Michael Osera} at the
University of Pennsylvania, who kindly shared their research project
and allowed me to run rampant in their codebase.  Their advice, insight,
and patience made this thesis possible. \\

\noindent\textbf{My parents and sister}, who endured the long
days and late nights that it took to bring this research to fruition and
kindly read countless
drafts of this seemingly incomprehensible paper.

%% file: proofs.tex
\section{Proofs}

\subsection{Notes}

In many of the judgments below, the constructor context (\CtorCtx) has been elided for the sake
of readability.  Since none of the synthesis or focusing judgments modify the constructor context,
it is assumed to be a fixed and globally available entity.

\subsection{Behavior of Focusing}


\subsubsection*{Lemma 2.1: Properties of Focusing}

\textsc{If} \hspace{1em}
\IFocus[
    \RefineCtx[\CtorCtx_1][\TypeCtx_1][\AuxCtx_1][\FocusCtx_1][\WorldCtx_1]
  ][\RefineCtx[\CtorCtx_2][\TypeCtx_2][\AuxCtx_2][\FocusCtx_2][\WorldCtx_2]
  ]
 \hspace{1em} \textsc{then} \hspace{1em}
$\CtorCtx_1 = \CtorCtx_2, \hspace{.5em}
 \TypeCtx_1 \subseteq \TypeCtx_2, \hspace{.5em}
 \AuxCtx_1 \subseteq \AuxCtx_2$

\vspace{.2cm}

\noindent \emph{Proof: Immediately follows from the focusing judgments. $\qed$}


\subsubsection*{Theorem 2.2: Progress of Focusing}

$\FocusCtx_1 = \cdot$ \hspace{1em} \textsc{or} \hspace{1em}
\IFocus[
    \RefineCtxM[\TypeCtx_1][\AuxCtx_1][\FocusCtx_1][\WorldCtx_1]
  ][\RefineCtxM[\TypeCtx_2][\AuxCtx_2][\FocusCtx_2][\WorldCtx_2]
  ]

\vspace{.2cm}

\noindent \emph{Proof: By case analysis on $\FocusCtx_1$.} \vspace{.1cm}\\
\noindent \emph{Case: $\FocusCtx_1 = \cdot$. $\qed$}       \vspace{.1cm}\\
\noindent \emph{Case: $\FocusCtx_1 = \FocusCtx,~\HasType[\Elim][\DType].$  Proof by case
                analysis on \DType.} \vspace{-.18cm}
\begin{tabbing}
\phantom{2em} \emph{Subcase: $\DType = \TUnit$.}  \hspace{3em} \=
              \emph{Apply \textsc{Focus-Unit}.} \= $\qed$
              \vspace{.1cm}\\
\phantom{2em} \emph{Subcase: $\DType = \TBase$.} \>
              \emph{Apply \textsc{Focus-Base}.} \> $\qed$
              \vspace{.1cm}\\
\phantom{2em} \emph{Subcase: $\DType = \TFunc{\DType_1}{\DType_2}$.} \>
              \emph{Apply \textsc{Focus-Fun}.} \> $\qed$
              \vspace{.1cm}\\
\phantom{2em} \emph{Subcase: $\DType = \TTuple{\DType_1 \TStar ... \TStar \DType_n}$}. \>
              \emph{Apply \textsc{Focus-Tuple}.} \> $\qed$
\end{tabbing}


\subsubsection*{Theorem 2.3: Preservation of Focusing\\}
\begin{tabbing}
\textsc{If} \hspace{1.5em} \=
               \Wf{\TypeCtx_1}, \hspace{.5em} \Wf{\AuxCtx_1},  \hspace{.5em} \Wf{\FocusCtx_1}
            \hspace{1em} \textsc{and} \hspace{1em} 
            \IFocus[
                    \RefineCtxM[\TypeCtx_1][\AuxCtx_1][\FocusCtx_1][\WorldCtx_1]
                  ][\RefineCtxM[\TypeCtx_2][\AuxCtx_2][\FocusCtx_2][\WorldCtx_2]]
            \\ \textsc{then} \>
            \Wf{\TypeCtx_2}, \hspace{.5em} \Wf{\AuxCtx_2},  \hspace{.5em} \Wf{\FocusCtx_2} 
\end{tabbing}
\vspace{.2cm}

\noindent \emph{Proof: By case analysis on the judgment 
               \IFocus[
                    \RefineCtxM[\TypeCtx_1][\AuxCtx_1][\FocusCtx_1][\WorldCtx_1]
                  ][\RefineCtxM[\TypeCtx_2][\AuxCtx_2][\FocusCtx_2][\WorldCtx_2]].}
          \vspace{.1cm}\\
\noindent \emph{Case: \textsc{Focus-Unit}.}

\begin{tabular}{l l l}
(1) & $\FocusCtx_1$ = $\FocusCtx_2, \HasType[\Elim][\TUnit]$ & By \textsc{Focus-Unit}\\
(2) & $\TypeCtx_2$  = $\TypeCtx_1, \HasType[\Elim][\TUnit]$ &  By \textsc{Focus-Unit}\\
(3) & $\AuxCtx_1 = \AuxCtx_2$ & By \textsc{Focus-Unit} \\
(4) & \Wf{\FocusCtx_1} & Assumption \\
(5) & \Wf{\AuxCtx_1} & Assumption \\
(6) & \Wf{\TypeCtx_1} & Assumption \\
(7) & \Wf{\FocusCtx_2} & 4 and inversion on \textsc{Type-Ctx-One-WF} \\
(8) & $\vdash \HasType[\Elim][\TUnit]$ & 4 and inversion on \textsc{Type-Ctx-One-WF} \\
(9) & \Wf{\AuxCtx_2} & (3) and (5) \\
(10)& \Wf{\TypeCtx_2} & (6), (9), \textsc{Type-Ctx-One-Wf} $\qed$ \\
\end{tabular}

\vspace{.1cm}
\noindent \emph{Case: \textsc{Focus-Base}, \textsc{Focus-Fun}. Similar. $\qed$} \vspace{.1cm}\\
\noindent \emph{Case: \textsc{Focus-Tuple}.}

\begin{tabular}{l l l}
(1) & $\FocusCtx_1$ = $\FocusCtx_2, \HasType[\Elim][\TTupleRange]$ & By \textsc{Focus-Tuple}\\
(2) & $\TypeCtx_2$  = $\TypeCtx_1$ &  By \textsc{Focus-Tuple}\\
(3) & $\AuxCtx_1 = \AuxCtx_2, \HasType[\EProj[1][\Elim]][\DType_1], ...,
                              \HasType[\EProj[n][\Elim]][\DType_n]$ & By \textsc{Focus-Tuple} \\
(4) & \Wf{\FocusCtx_1} & Assumption \\
(5) & \Wf{\AuxCtx_1} & Assumption \\
(6) & \Wf{\TypeCtx_1} & Assumption \\
(7) & \Wf{\FocusCtx_2} & 4 and inversion on \textsc{Type-Ctx-One-WF} \\
(8) & $\vdash \HasType[\Elim][\TTupleRange]$ & 4 and inversion on \textsc{Type-Ctx-One-WF} \\
(9) & $\forall i \in n,~\vdash \HasType[\EProj[i][\Elim]][\DType_i]$
    & (8) and \textsc{T-Proj} \\
(10)& \Wf{\AuxCtx_2} & (5), (9), and \textsc{Type-Ctx-One-WF} \\
(11)& \Wf{\TypeCtx_2} & (2) and (6) $\qed$ \\
\end{tabular}


\subsubsection*{Lemma 2.4: Potential of Focusing}

\noindent \emph{Let $\phi(\DType)$ be the size of the tree comprising the type \DType.}
          \vspace{.1cm}\\
\noindent \emph{Let $\Phi(\FocusCtx) = \sum_{\HasType[\Elim_i][\DType_i] \in \FocusCtx} \phi(\DType_i).$}
          \vspace{.1cm}\\
\noindent \textsc{If} \hspace{1em}
\IFocus[
    \RefineCtxM[\TypeCtx_1][\AuxCtx_1][\FocusCtx_1][\WorldCtx_1]
  ][\RefineCtxM[\TypeCtx_2][\AuxCtx_2][\FocusCtx_2][\WorldCtx_2]
  ]
 \hspace{1em} \textsc{then} \hspace{1em}
$\Phi(\FocusCtx_2) < \Phi(\FocusCtx_1)$

\vspace{.2cm}

\noindent \emph{Proof: By case analysis on the judgment \IFocus[
    \RefineCtxM[\TypeCtx_1][\AuxCtx_1][\FocusCtx_1][\WorldCtx_1]
  ][\RefineCtxM[\TypeCtx_2][\AuxCtx_2][\FocusCtx_2][\WorldCtx_2]
  ].}\vspace{.1cm}\\
\noindent \emph{Case: \textsc{Focus-Unit}.  $\FocusCtx_2 \subset \FocusCtx_1$, so
          $\Phi(\FocusCtx_2) < \Phi(\FocusCtx_1)$. $\qed$} \vspace{.1cm}\\
\noindent \emph{Case: \textsc{Focus-Base}, \textsc{Focus-Fun}. Similar. $\qed$} \vspace{.1cm}\\
\noindent \emph{Case: \textsc{Focus-Tuple.} Removes \HasType[\Elim][\TTupleRange] from $\FocusCtx_1$
                and adds \HasType[\EProj[1][\Elim]][\DType_1], ..., \HasType[\EProj[n][\Elim]][\DType_n]
                to $\FocusCtx_1$.  $\sum_{i \in n} \phi(\DType_i) < \phi(\TTupleRange)$, implying that
                $\Phi(\FocusCtx_2) < \Phi(\FocusCtx_1)$.} $\qed$


\subsubsection*{Theorem 2.5: Termination of Focusing}

\IFocusStar[
    \RefineCtxM[\TypeCtx_1][\AuxCtx_1][\FocusCtx_1][\WorldCtx_1]
  ][\RefineCtxM[\TypeCtx_2][\AuxCtx_2][\cdot][\WorldCtx_2]
  ]

\vspace{.2cm}

\noindent \emph{Proof: By the potential method and case analyis on $\FocusCtx_1$.} \vspace{.1cm}\\
\noindent \emph{Case: $\FocusCtx_1 = \cdot$.} Apply \textsc{Focus-Closure-Base}. $\qed$ \vspace{.1cm}\\
\noindent \emph{Case: $\FocusCtx_1 = \FocusCtx, \HasType[\Elim][\DType]$. By Theorem 2.2, we can
                apply a focusing judgment to any non-empty \FocusCtx.  Therefore, apply
                \textsc{Focus-Closure-Step}. By Lemma 2.4, $\Phi(\FocusCtx)$ decreases monotonically
                at each such step.  $\Phi(\cdot) = 0$, so this process eventually terminates
                with $\FocusCtx_1 = \cdot$.}

\subsection{Order of Focusing in the Proof Search Process}


\subsubsection*{Theorem 3.1: Focusing is Deterministic}

\begin{tabbing}
\textsc{If} \hspace{2em} \=
\IFocusStar[
    \RefineCtxM[\TypeCtx_1][\AuxCtx_1][\FocusCtx_1][\WorldCtx_1]
  ][\RefineCtxM[\TypeCtx_{2a}][\AuxCtx_{2a}][\cdot][\WorldCtx_2]
  ]
    \hspace{2em} \textsc{and} \hspace{2em}
\IFocusStar[
    \RefineCtxM[\TypeCtx_1][\AuxCtx_1][\FocusCtx_1][\WorldCtx_1]
  ][\RefineCtxM[\TypeCtx_{2b}][\AuxCtx_{2b}][\cdot][\WorldCtx_2]
  ]
    \\ \textsc{then} \>
$\TypeCtx_{2a} = \TypeCtx_{2b}$ \hspace{2em} \textsc{and} \hspace{2em}
$\AuxCtx_{2a} = \AuxCtx_{2b}$
\end{tabbing}
\noindent \emph{Proof: Immediately follows from the focusing judgments.} $\qed$

\subsubsection*{Theorem 3.2: Focusing Need Not Occur Before Synthesis}

\textsc{Define the judgments} \hspace{1em} \EGuessN
\hspace{1em} \textsc{and} \hspace{1em} \RefineN \hspace{1em}
\textsc{to be identical to those in Figure \ref{fig:productsrules} except that \FocusCtx~need not
be empty for any synthesis rule to be applied.}

\begin{tabbing}
\RefineM \= \hspace{2em} \textsc{iff} \hspace{2em} \RefineN \\
\EGuessM \> \hspace{2em} \textsc{iff} \hspace{2em} \EGuessN 
\end{tabbing}
\noindent \emph{Proof: Rather than focusing eagerly, we merely focus lazily as focused terms become
  necessary.  Since focusing is deterministic (Theorem 3.1), the time at which we perform focusing
  does not affect the expressions we can synthesize.} $\qed$

\subsection{Admissibility of Focusing}

\textsc{Define the judgments} \hspace{1em} \EGuessOM[\TypeCtx][\DType][\Elim]
\hspace{1em} \textsc{and} \hspace{1em} \RefineOM \hspace{1em}
\textsc{to be identical to those in Figure \ref{fig:productsrules} without focusing. Instead of focusing
to project on tuples, we introduce the EGuess-Proj rule as below:}
\begin{mathpar}
\inferrule*[right=EGuess-Proj]{
   \EGuessOM[\TypeCtx, \HasType[\Elim_1][\TTuple{\DType_1 \TStar ... \TStar \DType_m},~
             \HasType[\EProj[1][\Elim_1]][\DType_1],~...,~\HasType[\EProj[m][\Elim_1]][\DType_m]]
           ][\DType][\Elim]
}{
   \EGuessOM[\TypeCtx, \HasType[\Elim_1][\TTuple{\DType_1 \TStar ... \TStar \DType_m}]][\DType][\Elim]
}
\end{mathpar}

\subsubsection*{Theorem 4.1: Soundness of Focusing}

\begin{tabbing}
\noindent \textsc{If} \hspace{2em}
\RefineM[\TypeCtx][\cdot][\cdot][\cdot][\DType][\Intro] \=
\hspace{2em} \textsc{then} \hspace{2em}
\RefineOM[\TypeCtx][\cdot][\DType][\Intro] \\
\noindent \textsc{If} \hspace{2em}
\EGuessM[\TypeCtx][\cdot][\cdot][\DType][\Intro] \>
\hspace{2em} \textsc{then} \hspace{2em}
\EGuessOM[\TypeCtx][\DType][\Intro]
\end{tabbing}

\noindent \emph{Proof: We can transform any derivation of a judgment in the system with focusing into
the system without focusing.}
\vspace{.1cm}\\
\noindent \emph{Where elimination forms of unit, base, or function type are
placed in \FocusCtx~and subsequently moved into \TypeCtx~using
\textsc{IRefine-Focus} and \textsc{EGuess-Focus}, simply remove the focusing judgments and move
the elimination forms directly into \TypeCtx.}
\vspace{.1cm}\\
\noindent \emph{Where elimination forms of tuple type are
placed in \FocusCtx~and projected upon into their non-tuple constituents using
\textsc{IRefine-Focus} and \textsc{EGuess-Focus}, insert uses of the \textsc{EGuess-Proj} judgment
that perform the same projections.} $\qed$

\subsubsection*{Theorem 4.2: Completeness of Focusing}

\begin{tabbing}
\noindent \textsc{If} \hspace{2em}
\RefineOM[\TypeCtx][\cdot][\DType][\Intro] \=
\hspace{2em} \textsc{then} \hspace{2em}
\RefineM[\TypeCtx][\cdot][\cdot][\cdot][\DType][\Intro] \\
\noindent \textsc{If} \hspace{2em}
\EGuessOM[\TypeCtx][\DType][\Intro] \>
\hspace{2em} \textsc{then} \hspace{2em}
\EGuessM[\TypeCtx][\cdot][\cdot][\DType][\Intro]
\end{tabbing}

\noindent \emph{Proof: We can transform any derivation of a judgment in the system without focusing into
the system with focusing.}
\vspace{.1cm}\\
\noindent \emph{Whenever an elimination form is inserted into \TypeCtx, instead insert it into
\FocusCtx~and perform focusing.}
\vspace{.1cm}\\
\noindent \emph{Elimination forms of unit, base, or function type may be used exactly as before aside
from additional bookkeeping to move them from \FocusCtx~to \TypeCtx.}
\vspace{.1cm}\\
\noindent \emph{Even in the system
without focusing, tuples must still be fully projected before they may be used as a result of
the restriction that we can only synthesize programs that are eta-long. \textsc{EGuess-Ctx}
only moves terms from the context to goal position at base type.  No elimination form
judgment other than \textsc{EGuess-Proj} may make use of tuples.  Therefore, any tuple
placed into the context in the system without focusing is either unused or fully projected.
Focusing merely performs this projection eagerly.} $\qed$

\subsection{Soundness of Synthesis Judgments}

\subsubsection*{Theorem 5.1: Soundness of Synthesis Judgments}

\begin{tabbing}
\textsc{If} \hspace{1em}
\RefineM[\TypeCtx][\AuxCtx][\FocusCtx][\Overbar{n}{\World}]
\hspace{1em}
\=
\textsc{then}
\hspace{1em}
\HasTypeCtx[\Vars[\TypeCtx, \AuxCtx, \FocusCtx]][\Intro][\DType]
\hspace{1em}
\textsc{and}
\hspace{1em}
$\forall i \in n,~\ExamCtx_i(\Intro) \SssStar \DExam_i$
\\
\textsc{If} \hspace{1em}
\EGuessM[\TypeCtx][\AuxCtx][\FocusCtx][\DType][\Elim]
\hspace{1em}
\>
\textsc{then}
\hspace{1em}
\HasTypeCtx[\Vars[\TypeCtx, \AuxCtx, \FocusCtx]][\Elim][\DType]
\end{tabbing}

\vspace{.2cm}
{\footnotesize
\noindent \emph{Proof: By simultaneous induction on the \textsc{IRefine} and
   \textsc{EGuess} judgments.} \vspace{.1cm}\\
\noindent \emph{Case \textsc{EGuess-Ctx}.}

\begin{tabular}{l l l}
(1) & $\HasType[\Elim][\DType] \in \TypeCtx$ & Inversion $\qed$ \\
\end{tabular} \vspace{.1cm}\\
\noindent \emph{Case \textsc{EGuess-App}.}

\begin{tabular}{l l l}
(1) & \EGuessM[\TypeCtx, \HasType[\Elim_1][\TFunc{\DType_1}{\DType_2}]
          ][\AuxCtx][\HasType[\EApp[\Elim_1][\Intro_1]][\DType_2]][\DType][\Elim] & Inversion \\
(2) & \HasTypeCtx[\Vars[\TypeCtx, \HasType[\Elim_1][\TFunc{\DType_1}{\DType_2}], \AuxCtx,
                              \HasType[\EApp[\Elim_1][\Intro_1]][\DType_2]]] & Inductive Hypothesis \\
(3) & \HasTypeCtx[\Vars[\TypeCtx, \HasType[\Elim_1][\TFunc{\DType_1}{\DType_2}], \AuxCtx]]
                               & Vars relation $\qed$ \\
\end{tabular} \vspace{.1cm}\\

\noindent \emph{Case \textsc{EGuess-Focus}.}

\begin{tabular}{l l l}
(1) & \Vars[\TypeCtx_1, \AuxCtx_1, \FocusCtx_1] = \Vars[\TypeCtx_2, \AuxCtx_2, \FocusCtx_2]
    & Immediate from focusing judgments \\
(2) & \EGuessM[\TypeCtx_2][\AuxCtx_2][\FocusCtx_2][\DType][\Elim] & Inversion \\
(3) & \HasTypeCtx[\Vars[\TypeCtx_2, \AuxCtx_2, \FocusCtx_2]][\Elim][\DType] & Inductive Hypothesis \\
(4) & \HasTypeCtx[\Vars[\TypeCtx_1, \AuxCtx_1, \FocusCtx_1]][\Elim][\DType] & (1) and (3) $\qed$ \\
\end{tabular} \vspace{.1cm}\\

\noindent \emph{Case \textsc{IRefine-Focus}.}

\begin{tabular}{l l l}
(1) & \Vars[\TypeCtx_1, \AuxCtx_1, \FocusCtx_1] = \Vars[\TypeCtx_2, \AuxCtx_2, \FocusCtx_2]
    & Immediate from focusing judgments \\
(2) & \RefineM[\TypeCtx_2][\AuxCtx_2][\FocusCtx_2][\WorldN][\DType][\Elim] & Inversion \\
(3) & \HasTypeCtx[\Vars[\TypeCtx_2, \AuxCtx_2, \FocusCtx_2]][\Elim][\DType] & Inductive Hypothesis \\
(4) & \HasTypeCtx[\Vars[\TypeCtx_1, \AuxCtx_1, \FocusCtx_1]][\Elim][\DType] & (1) and (3) \\
(5) & $\forall i \in n,~\ExamCtx_i(\Intro) \SssStar \DExam_i$ & Inversion $\qed$ \\
\end{tabular} \vspace{.1cm}\\

\noindent \emph{Case \textsc{IRefine-Guess}.}

\begin{tabular}{l l l}
(1) & \EGuessM[\TypeCtx][\AuxCtx][\cdot][\TBase][\Elim]
    & Inversion \\
(2) & \HasTypeCtx[\Vars[\TypeCtx, \AuxCtx, \FocusCtx]][\Elim][\TBase] & Inductive Hypothesis \\
(3) & $\forall i \in n,~ \ExamCtx_i(\Elim) \SssStar \Exam_i$ & Inversion $\qed$ \\
\end{tabular} \vspace{.1cm}\\

\noindent \emph{Case \textsc{IRefine-Unit}.}

\begin{tabular}{l l l}
(1) & \HasTypeCtx[\Vars[\TypeCtx, \AuxCtx, \FocusCtx]][\EUnit][\TUnit] & \textsc{T-Unit} \\
(2) & $\forall i \in n,~ \ExamCtx_i(\EUnit) \SssStar \EUnit$ & \textsc{S-Closure-Base} $\qed$ \\
\end{tabular} \vspace{.1cm}\\

\noindent \emph{Case \textsc{IRefine-Ctor}.}

\begin{tabular}{l l l}
(1) & $\HasType[\DCtor][\TFunc{\DType}{\TBase}] \in \CtorCtx$
    & Inversion \\
(2) & \RefineM[\TypeCtx][\AuxCtx][\cdot
            ][\Overbar{i \in n}{\World[\ExamCtx_i][\Intro_i]}][\DType][\Intro] & Inversion \\
(3) & \HasTypeCtx[\Vars[\TypeCtx, \AuxCtx, \FocusCtx]][\Intro][\DType] & Inductive Hypothesis \\
(4) & \HasTypeCtx[\Vars[\TypeCtx, \AuxCtx, \FocusCtx]][\EApp[\DCtor][\Intro]][\TBase]
    & \textsc{T-Ctor}, (1), (3)\\
(5) & $\forall i \in n,~ \ExamCtx_i(\Intro) \SssStar \Intro_i$ & Inductive Hypothesis \\
(6) & $\forall i \in n,~ \ExamCtx_i(\EApp[\DCtor][\Intro]) \SssStar \EApp[\DCtor][\Intro_i]$
    & \textsc{S-Ctor}, (5) $\qed$ \\
\end{tabular} \vspace{.1cm}\\

\noindent \emph{Case \textsc{IRefine-Tuple}.}

\begin{tabular}{l l l}
(1) & $\forall i \in m,~ \RefineM[\TypeCtx][\AuxCtx][\cdot
                            ][\Overbar{j \in n}{\World[\ExamCtx_j][\Intro_{(i, j)}]}
                            ][\DType_i][\Intro_i]$ & Inversion \\
(2) & $\forall i \in m, \HasTypeCtx[\Vars[\TypeCtx, \AuxCtx, \FocusCtx]][\Intro_i][\DType_i]$
    & Inductive Hypothesis \\
(3) & \HasTypeCtx[\Vars[\TypeCtx, \AuxCtx, \FocusCtx]][\ETupleRange{\Intro}][\TTupleRange]
    & \textsc{T-Tuple, (2)} \\
(4) & $\forall (i, j) \in (m, n),~\ExamCtx_j(\Intro_i) \SssStar \Intro_{(i, j)}$
    & Inductive Hypothesis \\
(5) & $\forall j \in n,~\ExamCtx_j(\ETupleRange{\Intro}) \SssStar (\Intro_{(1, j)}, ..., \Intro_{(m, j)})$
    & \textsc{S-Tuple}, (4) $\qed$ \\
\end{tabular} \vspace{.1cm}\\

\noindent \emph{Case \textsc{IRefine-Match}.}

\begin{tabular}{l l l}
(1) & \EGuessM[\TypeCtx][\AuxCtx][\cdot][\TBase][\Elim] & Inversion \\
(2) & $\forall j \in m, ~\HasType[\DCtor_j][\TFunc{\DType_j}{\TBase}] \in \CtorCtx$ & Inversion \\
(3) & $\forall j \in m, ~\Refine[\CtorCtx][\TypeCtx][\AuxCtx][\HasType[\DVar][\DType_j]
                         ][\{\World[\ExamCtx_i, \Refines[\DVar][\DVal]][\Exam_i]~\big|~
                           \ExamCtx_i(\Elim) \SssStar \DCtor_j~\DVal\}
                         ][\DType][\Intro_j]$ & Inversion \\
(4) & \HasTypeCtx[\Vars[\TypeCtx, \AuxCtx, \FocusCtx]][\Elim][\TBase] & Inductive Hypothesis, (1) \\
(5) & $\forall j \in m,~
       \HasTypeCtx[\Vars[\TypeCtx, \AuxCtx, \FocusCtx, \HasType[x][\DType_j]]][\Intro_j][\DType]$
    & Inductive Hypothesis, (2) \\
(6) & \HasTypeCtx[\Vars[\TypeCtx, \AuxCtx, \FocusCtx]
                ][\EMatch[\Elim][\Overbar{j \in m}{\DCtor_j~\DVar \rightarrow \Intro_j}]][\DType]
    & \textsc{T-Match}, (2), (4), (5) \\
(7) & $\forall i \in n,~\ExamCtx_i(\Elim) \SssStar \EApp[\DCtor_j][\DVal]$ & Inversion \\
(8) & $\forall i \in n,~\textsc{if}~\ExamCtx_i(\Elim) \SssStar \EApp[\DCtor_j][\DVal]
                       ~\textsc{then}~\ExamCtx_i, \Refines[x][\DVal](\Intro_j) \SssStar \Exam_i$
    & Inductive Hypothesis, (3) \\
(9) & $\forall i \in n,~\ExamCtx_i(\EMatch[\Elim][\Overbar{j \in m}{\DCtor_j~\DVar \rightarrow \Intro_j}]) \SssStar \Exam_i$ & \textsc{S-Match1}, \textsc{S-Match2}, (7), (8) $\qed$ \\
\end{tabular} \vspace{.1cm}\\

\noindent \emph{Case \textsc{IRefine-Fix}.}

\begin{tabular}{l l l}
(1) &     \RefineM[\TypeCtx
          ][\AuxCtx][\HasType[\DFun][\TFunc{\DType_1}{\DType_2}], \HasType[\DVar][\DType_1]
          ][\Overbar{(i, k) \in (n, m_i)}{\World[\ExamCtx_i,
                                             \Refines[f][\Overbar{j \in m_i}{\DVal_{(i, j)}
                                                         \Rightarrow \DExam_{(i, j)}}],
                                             \Refines[x][\DVal_{(i, k)}]
                                           ][\DExam_{(i, k)}]}
          ][\DType_2
          ][\Intro] & Inversion \\
(2) & \HasTypeCtx[\Vars[\TypeCtx, \AuxCtx, \FocusCtx,
                        \HasType[\DFun][\TFunc{\DType_1}{\DType_2}], \HasType[\DVar][\DType_1]]
                ][\Intro][\DType_2] & Inductive Hypothesis \\
(3) & \HasTypeCtx[\Vars[\TypeCtx, \AuxCtx, \FocusCtx]
                ][\EFix[f][x][\DType_1][\DType_2][\Intro]
                ][\TFunc{\DType_1}{\DType_2}] & \textsc{T-Abs}, (2) \\
(4) & $\forall (i, k) \in (n, m_i),~\ExamCtx_i,
                                             \Refines[f][\Overbar{j \in m_i}{\DVal_{(i, j)}
                                                         \Rightarrow \DExam_{(i, j)}}],
                                             \Refines[x][\DVal_{(i, k)}](\Intro) \SssStar
                                             \DExam_{(i, k)}$ & Inductive Hypothesis \\
(5) & $\forall i \in n,~\ExamCtx_i(\EFix[f][x][\DType_1][\DType_2][\Intro]) \SssStar
        \Overbar{j \in m_i}{\DVal_{(i, j)} \Rightarrow \DExam_{(i, j)}}$ &
    \textsc{S-Closure-Base}, (4) $\qed$ \\
\end{tabular} \vspace{.1cm}\\ }

\noindent \emph{Note that we assume that the fixpoint and the partial function are equal to one another
  since they have the same behavior on all inputs for which the partial function is defined.}